\documentclass{aa}  

\usepackage{graphicx}
\usepackage{subcaption}
\usepackage{txfonts,textcomp}
\usepackage{hyperref}
\newcommand{\orcidlink}[1]{\protect\href{https://orcid.org/#1}{\protect\includegraphics[width=8pt]{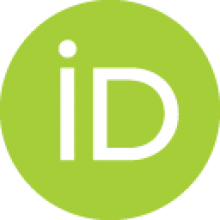}}}

\begin{document}

   \title{Mode conversion and energy flux absorption in the structured solar atmosphere}


   \author{S. J. Skirvin
          \inst{1}\orcidlink{0000-0002-3814-4232}
          \and
          T. Van Doorsselaere\inst{1}\orcidlink{0000-0001-9628-4113}
          }

   \institute{Centre for mathematical Plasma Astrophysics, Department of Mathematics, KU Leuven, Celestijnenlaan 200B bus 2400, B-3001 Leuven, Belgium\\
              \email{samuel.skirvin@kuleuven.be}
             }

   \date{Received --; accepted --}

 
  \abstract
   {Structuring in the solar atmosphere, in the form of inhomogeneities transverse to the magnetic field, is believed to play a vital role in wave propagation, conversion, and absorption.}
   {We investigated the effect of transverse structuring on the processes of mode conversion and wave energy flux absorption using a 3D ideal magnetohydrodynamic simulation featuring an expanding coronal loop in a gravitationally stratified atmosphere.}
   {Multiple wave drivers were modelled. The location of the driver at the photospheric base was allowed to vary so that we could study how the driven waves interact with the transverse structuring, provided by the magnetic field, as well as with the vertical structuring due to gravity.}
   {We find that the transverse structuring acts as a conduit for Alfv\'{e}n wave energy flux through the transition region and into the solar corona. Moreover, in regions of strong transverse gradients, the reflection of Alfv\'{e}n waves at the transition region is greatly reduced, supporting results from recent studies. Finally, we investigated the efficiency of the loop structuring at absorbing energy flux from externally driven waves and find that the loop is extremely effective at channelling wave energy flux to the loop apex in the corona; in some cases, it can absorb over a third of the externally driven wave energy flux.}
   {These results may have important consequences in the context of decayless loop oscillations as they suggest that the oscillations are driven by acoustic waves outside of the existing loop structure.}


   \maketitle
%

\section{Introduction} \label{sec:intro}

The solar coronal heating problem is still one of the fundamental unsolved problems in astrophysics \citep{vanDoorsselaere2020SSRv}. The feasibility of magnetohydrodynamic (MHD) waves providing a substantial contribution to the energy budget of the solar atmosphere has garnered increasing scientific interest over recent decades following the discovery of ubiquitous MHD waves throughout all layers of the Sun’s atmosphere \citep[e.g.][]{nak1999,McIntosh_et_al_2011,Jafarzadeh2017,morton2021,petrova2023, Yuan2023NatAs}. In particular, the detection of transverse motions in the solar corona with oscillation periods matching those of the internal pressure-driven acoustic resonances of the solar interior \citep{Oka2007,Tomczyk_et_al_2007, Anf2015, Morton2019, Gao2022} has reinforced the idea that MHD waves play a vital role in coupling and supplying energy throughout the solar atmosphere, and that they may even be the dominating contributing factor to coronal heating \citep{lim2023}. 

Recently, \citet{Skirvin2023ApJ} demonstrated that transverse oscillations of coronal loops, namely kink or Alfv\'{e}nic waves, with periods matching those from observations can be produced as a result of acoustic wave drivers mimicking those of solar p-modes. This is possible because they break the azimuthal symmetry of the system and as such the wave driver and the loop geometry are asymmetric. However, despite the presence of gravitational stratification in the model of \citet{Skirvin2023ApJ}, they did not provide a detailed discussion of the physical energy and mode conversion processes occurring in the lower atmosphere due to the driven acoustic waves that results in transverse motions in the corona with a magnetic nature.

The concept of mode conversion has an extensive history in the literature \citep[see e.g.][]{spruit1992, Schunker2006, Cally2008, cally2011, Khomenko2011, Khomenko2012, Felipe2012, Hansen2012, cally2019, Khomenko2019} and offers a promising pathway for acoustic energy, generated by convective motions in the solar interior, to develop magnetic properties and navigate the strongly stratified lower atmosphere, ultimately permeating the corona in the form of Alfv\'{e}nic waves \citep{morton2023}. 
Recently, the influence of transverse structuring on mode conversion has been studied in both an analytical and numerical context \citep{Cally2017, Khomenko2019}. These studies have focused on the fast magnetoacoustic-to-Alfv\'{e}n wave conversion in a domain with transverse structuring, with a particular interest in the effect of partial ionisation, which allows the Alfv\'{e}n wave to dissipate its energy through the process of ambipolar diffusion, and its important in the solar chromosphere. Ultimately, it was found that the transverse structuring plays a significant role in channelling Alfv\'{e}n waves into the corona, enhancing the power of the magnetic Poynting flux at greater heights, and reducing the reflection of Alfv\'{e}n waves at the transition region.

The transverse oscillations of a system of multiple loops or strands have been studied before \citep[e.g.][]{VanD2008loopsystem, Luna2008, Luna2010, TomVD2014, Rud2023}, with a particular focus on finding the eigenmodes of such a system. However, little attention has been paid to the efficiency of these loops to absorb wave energy from external oscillations, especially from acoustic waves, with some studies of simplified sunspot atmospheres suggesting that transverse structuring allows increased magnetic flux absorption \citep{Keppens1994}. Given how the turbulent solar surface is constantly generating waves through granular buffeting \citep[e.g.][]{Spruit1981, Choudhuri1993, Musielak2002, Stang2014}, it is to be expected that these waves will eventually interact with transverse inhomogeneities, in the form of, for example, magnetic elements, coronal loops, pores, and sunspots. Therefore, it is vital to understand how these potential wave guides absorb the energy flux from externally driven waves and ultimately channel this wave energy to the upper atmosphere.  

In this work we conduct a numerical investigation into the mode conversion processes resulting from acoustic wave drivers in a transversely structured and vertically stratified solar atmosphere. The questions that we wish to address are the following: (i) what role transverse structuring plays in the mode conversion of acoustic waves in the lower solar atmosphere, (ii) how the transverse structuring facilitates the channelling of different wave energy fluxes through the solar atmosphere, and (iii) how efficient loop-like structures are at absorbing the externally driven wave energy flux. 

The paper is structured as follows. In Sect. \ref{sec:methods} we introduce the numerical model and wave drivers for the simulations along with the wave mode decomposition approach. Our results on the mode conversion and absorption in the simulation are presented in Sect. \ref{sec:results} along with a discussion on the wave energy fluxes. Finally, we present a discussion and the conclusions of our results in Sect. \ref{sec:conclusions}.

\section{Methods} \label{sec:methods}
\subsection{Model} \label{subsec:model}
\begin{figure}
    \centering    \includegraphics[width=0.45\textwidth]{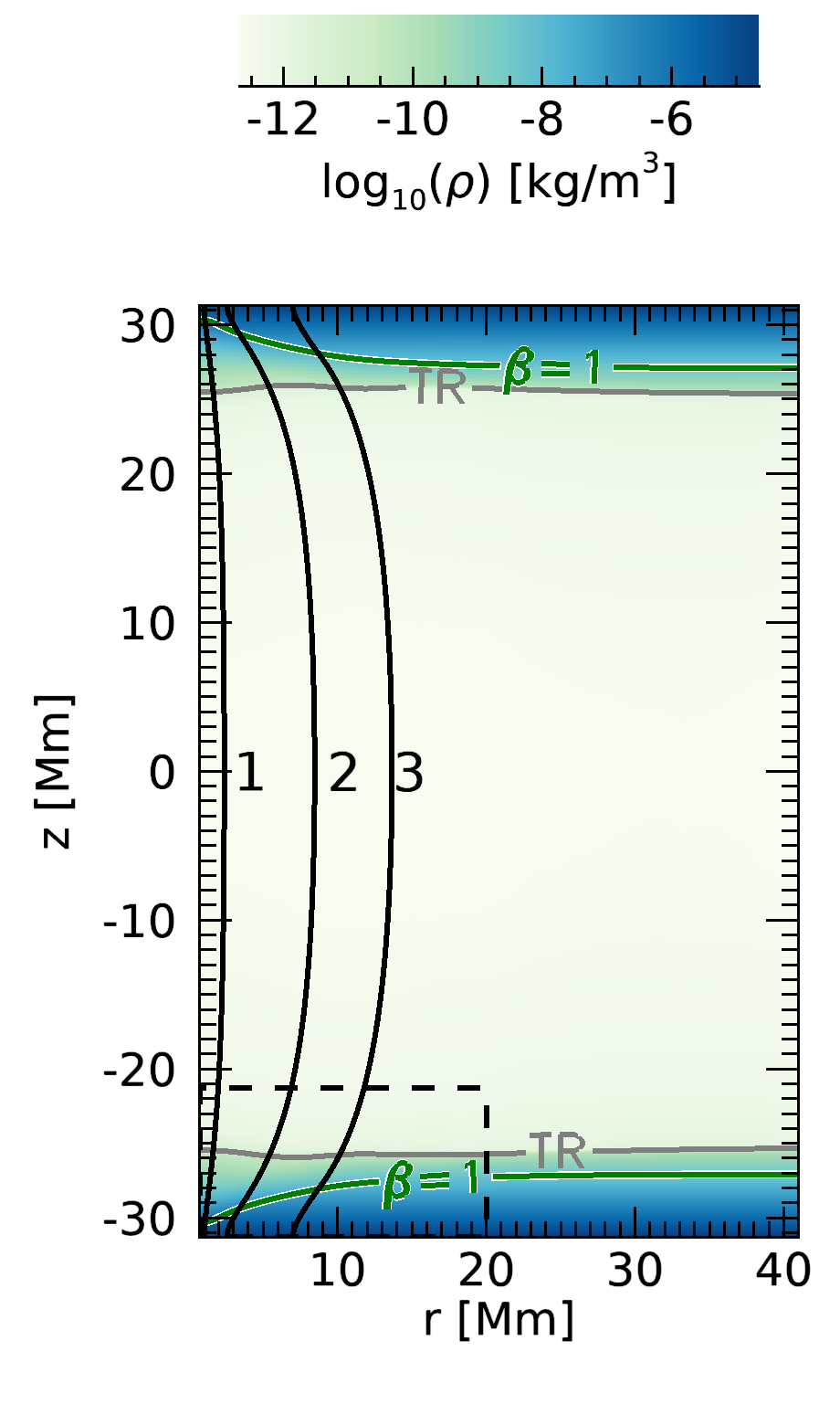}
    \caption{Snapshot of the initial background density (log scale) in the simulation domain. The full domain is plotted, ranging from photosphere to photosphere at the top and bottom boundaries, with the apex of the loop positioned at $z=0$ Mm. The positions of field lines $1$, $2,$ and $3$ are also indicated and used for the discussion in the text. Contours for the plasma-$\beta = 1$ layer and transition region are denoted by the green and grey lines, respectively. The black dashed box outlines the sub-domain used for analysis at various points in the present study.}
    \label{fig:background_rho}
\end{figure}

The model adopted in this work considers a coronal loop-like structure modelled as a local enhancement in the magnetic field strength and has been utilised in previous studies \citep[see e.g.][]{Reale2016,Riedl2021,Skirvin2023ApJ}. The model features a straightened, evacuated loop spanning from photosphere to photosphere at the top and bottom boundaries of the domain \citep{Guar2014} in a cylindrical coordinate system ($r,\varphi,z$). The background hydrostatic model is adapted from the work of \citet{Serio1981} for closed coronal loop models and the plasma density in the full domain is shown in Fig. \ref{fig:background_rho}. 
Additionally, we indicate selected separate field lines, $1$, $2,$ and $3,$ which will accompany the discussion in Sect. \ref{subsec:wave_fluxes}. Field line $1$ is rooted within the intense photospheric magnetic field, field line $2$ is rooted at the full width at half maximum of the Gaussian profile of the total magnetic field strength whereas field line $3$ is rooted at the centre of the locally applied shifted wave drivers (see Sect. \ref{subsec:drivers}), positioned at $r=7$ Mm. A vertical magnetic field is then introduced into the atmosphere with a Gaussian profile in strength \citep{Reale2016}, with a magnitude of $273$ G on the axis, which reduces to around $10$ G in the ambient atmosphere. We note that in our model the transverse structuring provided by the magnetic field is a smooth function with stronger gradients at the loop apex, unlike recent work by \citet{Khomenko2019} where the transverse structuring is provided through a repetition of small-scale flux tubes. The atmosphere is then allowed to relax numerically until the background velocities become negligible. As a result of the numerical relaxation process, the magnetic field expands in the corona to maintain total pressure balance, at which the field strength on the axis at the apex of the loop is $13$ G. This numerically relaxed atmosphere with the expanded magnetic field enhancement is taken as the equilibrium configuration for the simulations; doing so, and by conducting a simulation both with and without a wave driver, we can calculate the perturbed quantities by subtracting the simulation without the wave driver from the simulation with the wave driver.
\begin{figure}
    \centering    \includegraphics[width=0.45\textwidth]{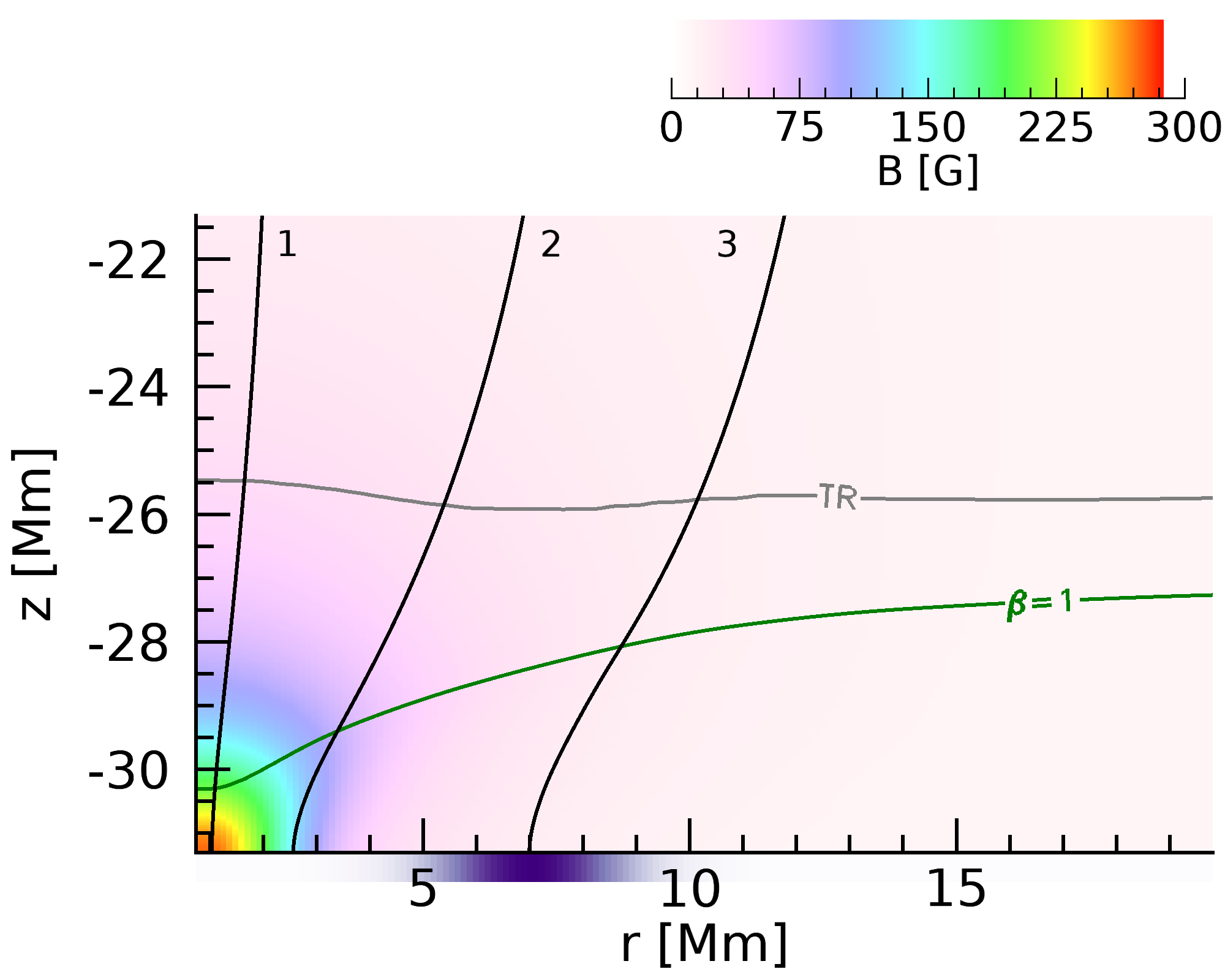}
    \caption{Snapshot of the initial magnetic field strength and topology. The same configuration is found for all azimuthal angles due to the axi-symmetry of the equilibrium. The location and strength of the shifted driver (used in both WD2 and WD3) is shown by the purple bar on the horizontal axis. The positions of field lines $1$, $2,$ and $3$ are also indicated and used for the discussion in the text. Contours for the plasma-$\beta = 1$ layer and transition region are denoted by the green and grey lines, respectively.}
    \label{fig:Bfield_strength}
\end{figure}
The initial background magnetic field strength in the domain is displayed in Fig. \ref{fig:Bfield_strength}. At the location of the shifted drivers outside of the loop, the magnetic field strength is essentially homogeneous in the transverse direction, with a value of around $10$ G; however, the key aspect to note here is that the transverse gradients are important on field line $1$ and the transverse gradients of the magnetic field strength are small in the vicinity of field line $3$. Furthermore, to aid understanding of wave propagation, we also display the characteristic speeds (Alfv\'{e}n and acoustic) in the lower atmosphere and their horizontal stratification in Fig. \ref{fig:characteristic_speeds}.

\subsection{Numerical setup and boundary conditions}\label{subsec:setup}

The numerical domain and grid cell size adopted in this work is the same as prescribed in \citet{Skirvin2023ApJ}. Our simulation domain ranges from $0.73$ Mm to $41.01$ Mm in the radial direction, from $0$ to $\pi$ in the $\varphi$ direction, and from $-31.31$ Mm to $31.31$ Mm in the $z$ direction, with $192$, × $256$, and × $768$ data points, respectively. The loop axis is located at $r=0;$ however, we did not simulate close to this region due to the regular singularity at the origin of the domain when using a polar cylindrical coordinate geometry. An exact description of the stretched numerical grid and resolution in different regimes can be found in \citet{Riedl2021} and \citet{Skirvin2023ApJ}.

The simulations were performed using the PLUTO code \citep{Mign2007}, where the ideal MHD equations were solved using the Harten-Lax-Van Leer approximate Riemann solver, with a piece-wise total variation diminishing linear reconstruction method for the spatial integration. We utilised reflective boundary conditions for both boundaries in the $r$ direction and anti-symmetric boundary conditions in the $\varphi$ direction, where the signs for the tangential components of the magnetic field and velocity field are reversed. Additionally, we incorporated anti-symmetric boundaries for the upper $z$ boundary. At the lower $z$ boundary, the same boundary conditions were set; however the velocity, pressure, and density were perturbed according to an analytical solution for a gravity-acoustic wave, given by the description in Sect. \ref{subsec:drivers}.

\subsection{Wave drivers} \label{subsec:drivers}
\begin{figure}
    \centering    \includegraphics[width=0.45\textwidth]{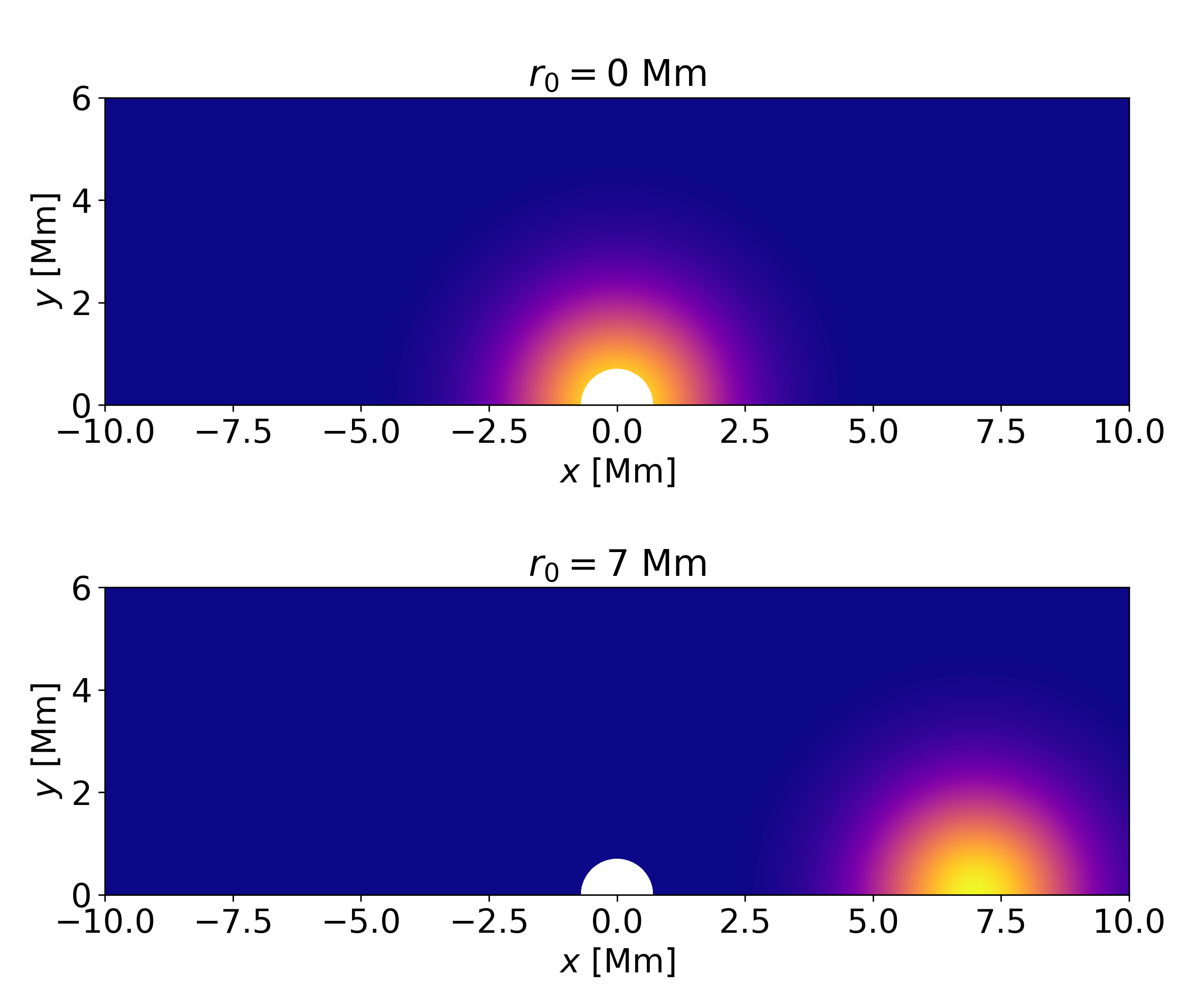}
    \caption{Location of the drivers with $r_0=0$ Mm and $r_0=7$ Mm employed in this study (displayed on a Cartesian grid). The strength of the driver is shown in arbitrary units. Note that the gap at the centre of the $x-y$ plane is due to the numerical domain starting at $r=0.71$ Mm in the radial direction.}
    \label{fig:driver_loc_2D}
\end{figure}

In this study we implemented three different wave drivers, WD1, WD2, and WD3, all with a Gaussian spatial dependence, in order to investigate the mode conversion and absorption in a transversely structured atmosphere. The only difference between the three wave drivers was the location of the perturbations and the angle of inclination, $\theta$, with respect to the vertical axis, $z$. A summary of the properties of the three wave drivers is provided in Table \ref{tab:wave_drivers} and a 2D crosscut at the base of the numerical domain is displayed in Fig. \ref{fig:driver_loc_2D} to give an idea of the location of the driver(s) employed in this work.

\begin{table*}
\caption{Properties of the three Gaussian wave drivers (WD1, WD2, and WD3) considered in this work. The properties are the same except for the spatial localisation (where the centre of the Gaussian is located at $r_0$) and inclination with respect to the vertical axis, $\theta$. WD2 and WD3 are located outside the loop structuring and will be used to study the effectiveness of the loop structuring at absorbing the wave energy flux. WD2 is a vertical wave driver whereas WD3 is inclined at an angle of $20^{\circ}$ towards the loop.}             
\label{tab:wave_drivers}      
\centering          
\begin{tabular}{c c c c c c} 
\hline\hline       
                     
Wave Driver (WD) & Amplitude [ms$^{-1}$] & Period [s] & Width [Mm] & $r_0$ [Mm] & $\theta$ [$^{\circ}$] \\ 
\hline                    
WD1 & 300 & 370 & 2 & 0.0 & 15 \\
WD2 & 300 & 370 & 2 & 7.0 & 0 \\
WD3 & 300 & 370 & 2 & 7.0 & -20 \\
\hline                  
\end{tabular}
\end{table*}

In all cases, the wave drivers were multiplied with the analytical solution for a gravity-acoustic wave, which is appropriate as our bottom boundary is located in the photosphere where the plasma-$\beta$ is much greater than unity. Therefore, we perturbed the velocity, plasma density, and plasma pressure according to \citep{Mihalis1984, Santamaria2015, Skirvin2023ApJ}
\begin{eqnarray}
    v_r' = &A& |v| \exp \left(\frac{z}{2H}+\Im(k_z)z\right)\cos(\varphi) \nonumber \\ &\times&\sin \left( \omega t - \Re(k_z)z -k_{\perp}r \cos(\varphi)+\phi_v \right), \label{eq:driver_velocityr}\\    
    v_{\varphi}'= &-A& |v| \exp \left(\frac{z}{2H}+\Im(k_z)z\right)\sin(\varphi) \nonumber \\ &\times&\sin \left( \omega t - \Re(k_z)z -k_{\perp}r \cos(\varphi)+\phi_v \right), \label{eq:driver_velocityphi}\\            
    v_z'= &A& \exp \left(\frac{z}{2H}+\Im(k_z)z\right) \nonumber \\ &\times&\sin \left( \omega t - \Re(k_z)z -k_{\perp}r \cos(\varphi)\right), \label{eq:driver_velocityz}\\
    p' = &A& |P| \exp \left(\frac{z}{2H}+\Im(k_z)z\right) \nonumber \\ &\times&\sin \left( \omega t - \Re(k_z)z -k_{\perp}r \cos(\varphi) + \phi_P\right), \label{eq:driver_pressure}\\
    \rho' = &A& |\rho| \exp \left(\frac{z}{2H}+\Im(k_z)z\right) \nonumber \\ &\times&\sin \left( \omega t - \Re(k_z)z -k_{\perp}r \cos(\varphi) + \phi_{\rho}\right), \label{eq:driver_density}
\end{eqnarray}
where detailed expressions for all the coefficients and terms in Eqs. (\ref{eq:driver_velocityr})-(\ref{eq:driver_density}) can be found in \citet{Skirvin2023ApJ} and remain unchanged in this work. The notation of $\Im(k_z)$ and $\Re(k_z)$ denotes the imaginary and real components of the vertical wavenumber of the driven waves. Here, A=$300$ m s$^{-1}$ is the driver amplitude, which agrees with observational photospheric Doppler oscillations from the contribution of p-modes \citep{McClure2019}, although an individual p-mode alone will possess a much smaller velocity amplitude \citep{Priest2014}. The relative amplitudes for the velocity, pressure and density perturbations are denoted as $|v|$, $|P|$ and $|\rho|$, respectively. Moreover, $H$ is the pressure scale height; $k_z$ is the vertical wavenumber, which only has a real part in our case; $\omega = 2\pi/T$ is the driver frequency, with period $T = 370$ s, which is within the typical range of p-mode periods. A detailed discussion regarding the gravitational cutoff frequency in comparison with the driving frequency can be found in \citet{Riedl2021}. The variables $\phi_v$, $\phi_P$, and $\phi_{\rho}$ are the velocity, pressure, and density phase shifts compared to the vertical velocity perturbation, $\hat{v}_z$. The perpendicular wavenumber $k_{\perp}=\sqrt{k_r^2+k_{\varphi}^2}$ of the driven waves (see Eqs. (13)-(16) in \citealt{Skirvin2023ApJ}) in this simulation is constant for all azimuthal angles and all times, although the radial structuring in the model and the localisation of the driver result in a horizontal wavenumber that depends on the radial position. A more detailed discussion on the calculation of the perpendicular wavenumber in the simulations is provided in Sect. \ref{subsec:mode_conversion}. The presence of the term $k_{\perp}$ arises when the driver is inclined with respect to the vertical axis of the domain and results in waves that are propagating in space in the plane orthogonal to the magnetic field.

Furthermore, the wave drivers implemented are localised in space. This is achieved by multiplying Eqs. (\ref{eq:driver_velocityr})-(\ref{eq:driver_density}) by the function
\begin{equation}
    D(r) = \exp\left( - \frac{\left(r \cos{\varphi}-r_0\right)^2}{\sigma_r^2} -  \frac{\left(r\sin{\varphi} - \varphi_0 \right)^2}{\sigma_{\varphi}^2}\right),
\end{equation}where $\sigma_{r,\varphi}$ is the standard deviation of the Gaussian describing the width of the driver, in both the radial and azimuthal directions, which in this study are taken to be $\sigma_r = \sigma_{\varphi} = 2$ Mm, resulting in an elliptical  Gaussian distribution in the $2$D plane at the base of the domain. Both $r_0$ and $\varphi_0$ have units of distance and represent the spatial `shift' of the driver in both the radial and azimuthal directions in the cylindrical domain. All wave drivers considered in this work are localised in the azimuthal direction by setting $\varphi_0 = 0$ such that the Gaussian is centred on azimuthal angle $\varphi = 0$. We note that by implementing a wave driver locally in both the radial and azimuthal directions, this results in driven waves with both a radial and azimuthal wavenumber. Therefore, even when considering a purely vertical wave driver, if the wave driver is localised in space with $r_0 \ne 0$, then kink/Alfv\'{e}nic motions can be produced such as those presented in \citet{Skirvin2023ApJ}. 

\subsection{Vector component decomposition}\label{subsec:decompositon}
Due to the expansion of the magnetic field in the chromosphere and corona, we do not have a purely vertical magnetic field in the model and it would be useful to decompose vector components respective to the background magnetic field vector in order to provide a discussion on the different wave modes present.

In full 3D simulations, where the magnetic field is not confined to a 2D geometry, the isolation of MHD waves becomes non-trivial as there are an infinite number of vectors normal to the magnetic field vector \citep{Yadav2022}. To help distinguish between the different types of waves in our simulation, we adopted a decomposition method similar to that used in \citet{Riedl2019}. The conversion of components from a cylindrical geometry ($r, \varphi, z$) to those parallel, normal, and tangential to magnetic flux surfaces is given by\begin{eqnarray}
    \mathbf{e}_{\parallel} &=& \left[\frac{B_r \text{cos}(\varphi)}{\sqrt{B_r^2 + B_z^2}}, \frac{B_r \text{sin}(\varphi)}{\sqrt{B_r^2 + B_z^2}}, \frac{B_z(\varphi)} {\sqrt{B_r^2 + B_z^2}}\right], \label{decomp_parallel} \\
    \mathbf{e}_{\phi} &=& [-\text{sin}(\varphi), \text{cos}(\varphi), 0], \label{decomp_azi} \\
    \bf{e}_{\perp} &=& \bf{e}_{\varphi} \times \bf{e}_{\parallel}, \label{decomp_perp}
\end{eqnarray}
where $\mathbf{e}$ denotes a unit vector in each direction, respectively. Equations (\ref{decomp_parallel})-(\ref{decomp_perp}) set up a Cartesian basis that describes the vector decomposition with respect to magnetic field lines for an equilibrium magnetic field that is structured in the $r$ and $z$ directions in a cylindrical geometry. The component of magnetic field azimuthal to magnetic surfaces is ignored in the decomposition due to the field lines being circularly symmetric around the axis of the loop and no magnetic twist is considered in the initial model. The formalism presented in Eqs. (\ref{decomp_parallel})-(\ref{decomp_perp}) provides a proxy to separate the slow and fast magnetoacoustic waves in a low beta plasma; however, the decomposition for the azimuthal component tangent to magnetic flux surfaces is valid for Alfv\'{e}n waves in a high beta plasma also. This decomposition of components parallel, perpendicular, and azimuthal to the magnetic field lines will be important in the context of understanding the wave modes that are present in the simulation.

\section{Results} \label{sec:results}

\subsection{Mode conversion and the effect of transverse structuring}\label{subsec:mode_conversion}

The study by \citet{Khomenko2012} \citep[see also][]{Felipe2012} found that acoustic waves in a sunspot atmosphere convert to magnetoacoustic modes at the equipartition layer. The incident (fast) acoustic wave can be partially converted to both a fast (magnetic) and slow (magnetic) wave depending on the attack angle. Importantly, the authors show that fast magnetoacoustic waves undergo total reflection in the vicinity of the fast mode reflection height and do not propagate higher than a few hundred kilometres into the sunspot atmosphere. This mode conversion process shares similarities with the resonant absorption mechanism in a transversely structured environment \citep{Cally2010}. Here, our aim is to investigate the effect of transverse structuring on mode conversion as the presence of a structured waveguide may support additional wave propagation.

\begin{figure*}
    \centering    \includegraphics[width=0.95\textwidth]{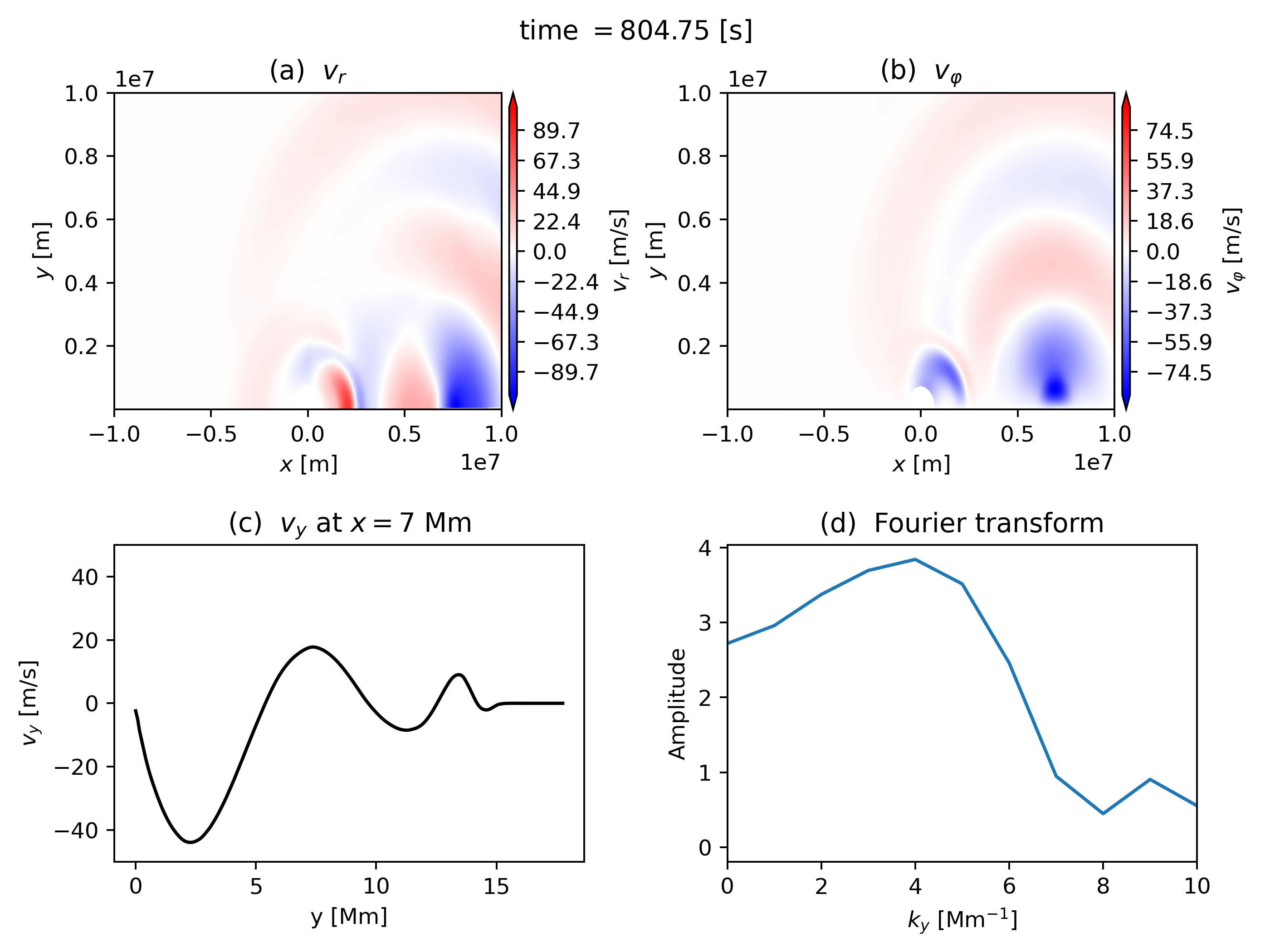}
    \caption{Snapshots of the radial (a) and azimuthal (b) velocity perturbations across the bottom boundary of the domain at a time $t=805$ s in the shifted wave driver simulation WD2. The signal for the velocity perturbation $v_y$ (after converting to Cartesian coordinates) is shown in panel (c) for a slice taken along $x=7$ Mm, where the shifted drivers (WD2 and WD3) are positioned. The resulting Fourier transform (d) of the signal in panel (c) allows us to compute the fast mode reflection height $z_{\text{refl}}$ at this moment in space and time.}
    \label{fig:fourier_kperp}
\end{figure*}

Firstly, it is instructive to provide a more detailed discussion on the fast mode reflection height ($z_{\text{refl}}$). In both studies by \citet{Khomenko2012} and \citet{Felipe2012}, the horizontal wavenumbers were taken to be a fixed value of $1.37$ Mm$^{-1}$. The fast mode reflection height can then be computed as the height $z_{\text{refl}}$ where $\omega=v_A k_{\perp}$. This is possible to determine analytically when considering driven monochromatic plane waves by perturbing the entire bottom boundary of the simulation domain. However, in our simulations, we applied a localised driver, which itself may be inclined by various angles with respect to the vertical axis of the domain. Both the inclination and the localisation of the driver contribute to the resulting value of horizontal wavenumber, $k_{\perp}$. Determining the exact value of $k_{\perp}$ is made even more difficult by the transverse inhomogeneity of the background model. Therefore, we opted to determine the horizontal wavenumber in our simulations numerically, as opposed to analytically. We did so by taking a Fourier transform of the perturbed velocity signal in the horizontal direction (e.g. $\hat{v}_r$ as a function of $r$) and locating the peak in power of the resulting frequency spectra $k_r$. An example of this is shown in Fig. \ref{fig:fourier_kperp} for the WD2 simulation at a specific snapshot in time. In Fig. \ref{fig:fourier_kperp} a spatial slice is taken at $x=7$ Mm (after converting to Cartesian coordinates) for the $v_y$ velocity perturbation to calculate the Fourier components, where, a peak is located at $k_{y} = 4$ Mm$^{-1}$. Finally, as the perturbations produce propagating circular wave fronts, the horizontal wavenumber $k_{\perp}$ can be considered equal to the radial wavenumber $k_r$ (or, in a Cartesian geometry $k_x$ or $k_y$). We find that the resulting horizontal wavenumbers in all simulations are time and space dependent and peak in the region of $1.5-6.0$ Mm$^{-1}$ corresponding to a reflection height (where $\omega=v_A k_{\perp}$) range spanning $1.5$ Mm in the chromospheric region of our model. These wavenumbers are also consistent with the values from the sunspot model of \citet{Khomenko2012} and \citet{Felipe2012} who considered a greater value of magnetic field strength (roughly three times larger at the axis). As a result of the localisation of the wave driver, waves with varying wavelengths are produced, ranging from roughly $0.14$ Mm to greater than $10$ Mm. Previous studies including \citet{Khomenko2019} utilised tube structuring, whereby the spatial wavelengths of the inhomogeneities were smaller than the driven wavelengths, unlike the present model of a smooth inhomogeneity with stronger gradients at the axis of the loop. Therefore, it is likely that in some spatial locations in our model, where the structuring and transverse gradients are stronger (namely closer to the loop axis), similar processes regarding mode conversion and absorption to those reported by \citep{Khomenko2019} are present. However, in other locations such as far away from the loop structuring, where the driven wavelengths are larger than the characteristic inhomogeneity spatial scale, the physical picture is likely different as the waves will not `experience' the structuring to the same degree.

Using the wave decomposition method presented in Sect. \ref{subsec:decompositon}, we could create a proxy to separate the waves in the simulation into slow magnetoacoustic, fast magnetoacoustic, and Alfv\'{e}n to allow a comparison with the results of \citet{Khomenko2012}; however, we stress that this formalism is only valid in a low-$\beta$ environment, with the exception of the decomposition for Alfv\'{e}n waves, which remains valid in all plasma-$\beta$ regimes. Therefore, in the following discussion, the decomposition of wave modes into slow, fast, and Alfv\'{e}n is only physically meaningful above the equipartition layer. For the purpose of this section, we scaled the velocity amplitudes by a factor of $\sqrt{\rho v_{ph}}$, where $v_{ph}=c_s$ for the slow magnetoacoustic mode ($\hat{v}_{\parallel}$), $v_{ph}=\sqrt{c_s^2+v_A^2}$ for the fast magnetoacoustic mode ($\hat{v}_{\perp}$), and $v_{ph}=v_A$ for the Alfv\'{e}n mode ($\hat{v}_{\phi}$). Scaling the velocities with this factor enabled us to study the energy flux that is propagating, in other words, it provides a proxy to investigate the energy that is transported by each wave mode. This formalism allows for a direct discussion of the similarities and differences between the scaled velocities in our structured 3D model with those displayed in Fig. 3 of \citet{Khomenko2012}. Similar to previous studies, above the layer where $c_s=v_A$, the velocity amplitude of the Alfv\'{e}n mode is still significantly smaller than that of the slow magnetoacoustic mode, due to the nature of the driver injecting perturbations mainly along the magnetic field.
\begin{figure*}
\begin{subfigure}{0.97\textwidth}
    \includegraphics[width=\textwidth]{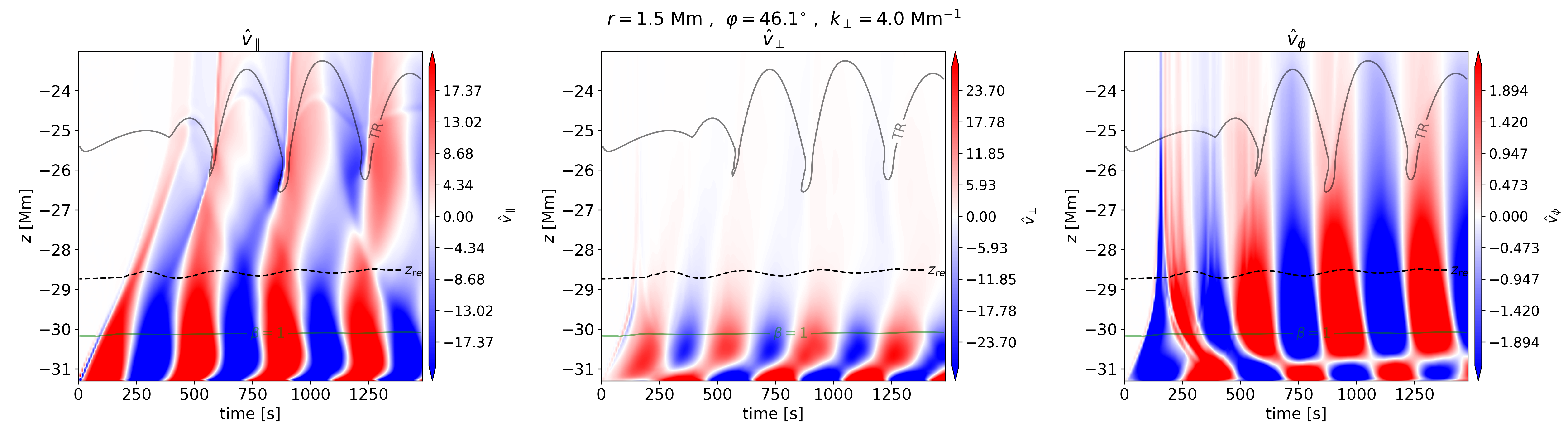}
    \caption{}
\end{subfigure}
\begin{subfigure}{0.97\textwidth}
    \includegraphics[width=\textwidth]{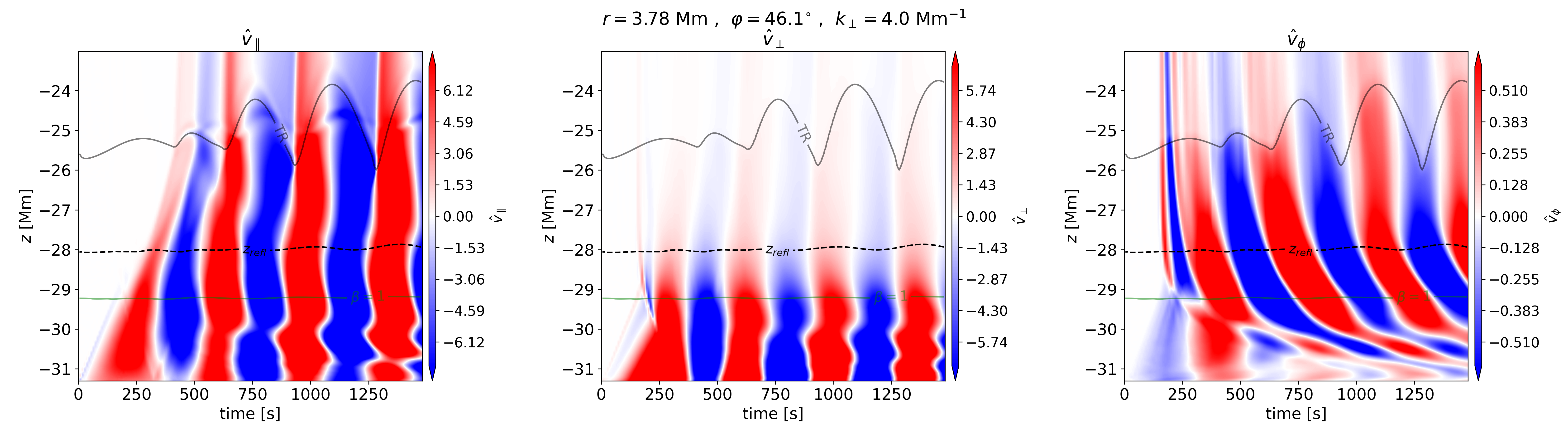}
    \caption{}
\end{subfigure}
\begin{subfigure}{0.97\textwidth}
    \includegraphics[width=\textwidth]{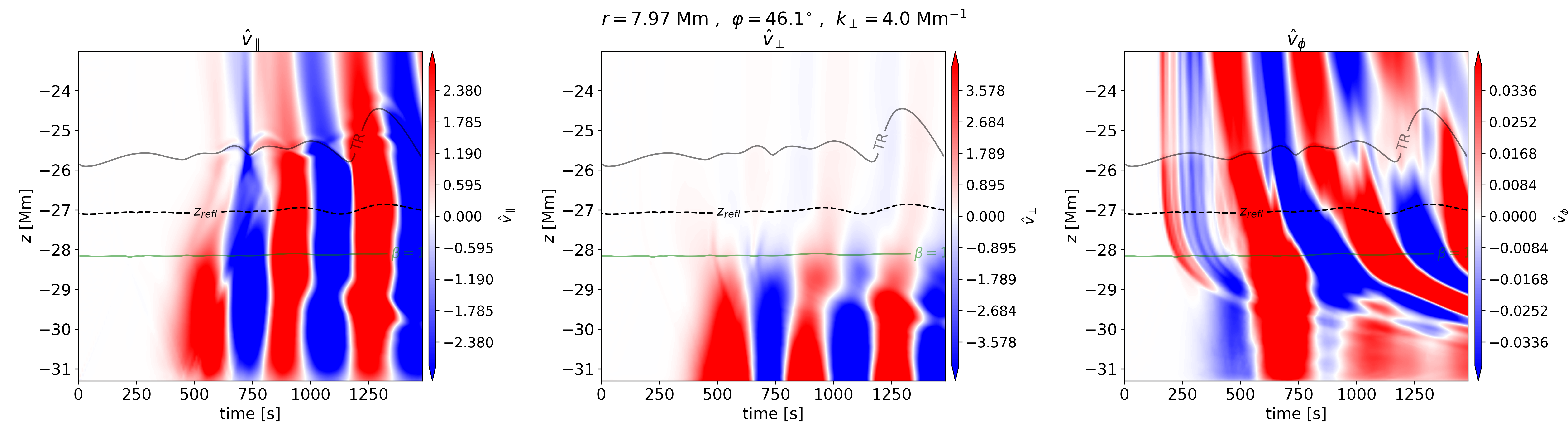}    
    \caption{}
\end{subfigure}
\caption{Time--distance plots of the scaled velocities for the perturbations parallel, perpendicular, and azimuthal to magnetic flux surfaces, a proxy for the slow magnetoacoustic mode (left), fast magnetoacoustic mode (middle)n and Alfv\'{e}n mode (right), respectively. All plots are for the case of a localised inclined driver of $15^{\circ}$ (WD1) for an azimuthal slice $\varphi=46.1^{\circ}$ at three different radial cuts of (a) $r=1.5$ Mm, (b) $r=3.78$ Mm and (c) $r=7.97$ Mm. The $\beta=1$, fast mode reflection ($z_{\text{refl}}$) and transition region layers are shown by the green, black dashed and grey contours, respectively. For the purpose of this illustration, the fast mode reflection height was calculated using a horizontal wavenumber $k_{\perp}= 4.0$ Mm$^{-1}$. Note the different colour bar used for the $\hat{v}_{\parallel}$, $\hat{v}_{\perp}$ and $\hat{v}_{\phi}$ components in all panels.
\label{fig:scaled_vels_15deg}}
\label{fig:scaled_vels_15deg}
\end{figure*}

The resulting time--distance diagrams of the scaled velocities are shown in Fig. \ref{fig:scaled_vels_15deg} for WD1: the case where the driver is centred on $r=0$ and inclined by $15^{\circ}$. This figure shows the analysis conducted at a fixed azimuthal angle $\varphi=46.1^{\circ}$ at different radii in the model. In all cases, it is evident that there is conversion from slow to fast magnetoacoustic around the equipartition layer (near $\beta=1$). The resulting time--distance diagram for the slow magnetoacoustic mode does not appear to vary much at different radii, the only difference being the phase shift of the wave fronts, which is entirely expected due to the curvature of the field lines at larger radii resulting in waves travelling longer distances along the field at greater radii. Moreover, paying attention to the fast magnetoacoustic mode, denoted by $\hat{v}_{\perp}$, there are strong similarities with the study of a thick flux tube sunspot model in \citet{Khomenko2012}. In addition, there is a good confirmation for our numerically determined horizontal wavenumber $k_{\perp}$ in the time--distance plot of the $\hat{v}_{\perp}$ component, as the scaled velocity $\hat{v}_{\perp}$ appears to decay above the determined fast mode reflection height $z_{\text{refl}}$ where 3D conversion to Alfv\'{e}n waves occurs.

Turning attention now to the Alfv\'{e}n mode shown in the right panels of Fig. \ref{fig:scaled_vels_15deg}. There is clear evidence of conversion from fast magnetoacoustic to Alfv\'{e}n around $z_{\text{refl}}$ as predicted by \citet{Khomenko2012}, this can be seen by the distinct wave fronts of the $\hat{v}_{\phi}$ component above $z_{\text{refl}}$, appearing almost standing in nature at small radii. There are also reflections of the Alfv\'{e}n wave from the transition region, which can be seen more clearly in Figs. \ref{fig:scaled_vels_15deg}b and \ref{fig:scaled_vels_15deg}c, and is a feature that has been known in solar physics for a long time \citep[e.g.][]{cra2005}. However, the presence of reflections is less pronounced at radii closer to the axis of the loop, where the field is more vertical, transverse gradients are stronger and the overall field strength is greater, when compared to larger radii where the field is weaker and more inclined. Instead, the seemingly constant phase speed of the Alfv\'{e}n waves above $z_{\text{refl}}$ in Fig. \ref{fig:scaled_vels_15deg}a hints that the loop structuring reduces the reflections from the transition region, this was also a major conclusion in the study of \citet{Khomenko2019}. In addition, the azimuthal perturbations associated with the global transverse motions of non-uniform flux tubes \citep{Goossens2020,Skirvin2022} due to the coupling between the propagating kink waves and Alfv\'{e}n waves \citep{Pascoe2011} may also contribute to the standing wave fronts in the $\hat{v}_{\phi}$ time--distance map. It is also quite possible that Alfv\'{e}n waves are generated in the simulation through the non-linear coupling to the slow waves \citep{hollweg1971,Kudoh1999,Ballester2020}, which display non-linear characteristics (see Fig. \ref{fig:scaled_vels_15deg}). However, we anticipated that the non-linear effects would be less dominant than the mode conversion processes, as the applied driver amplitude is less than the background velocities in the simulation domain.

To study the effect that the transverse structuring has on mode conversion and wave absorption (Fourier scattering), we repeated the same analysis but using a setup where the wave driver is positioned outside the loop structuring, in a vicinity where the transverse gradients are much weaker. We centred the localised wave driver at $r_0=7$ Mm (see Table \ref{tab:wave_drivers}) and considered both a vertical driver (WD2) and a driver inclined towards the loop (WD3) with an angle of $20^{\circ}$. 

\begin{figure*}
\begin{subfigure}{0.97\textwidth}
    \includegraphics[width=\textwidth]{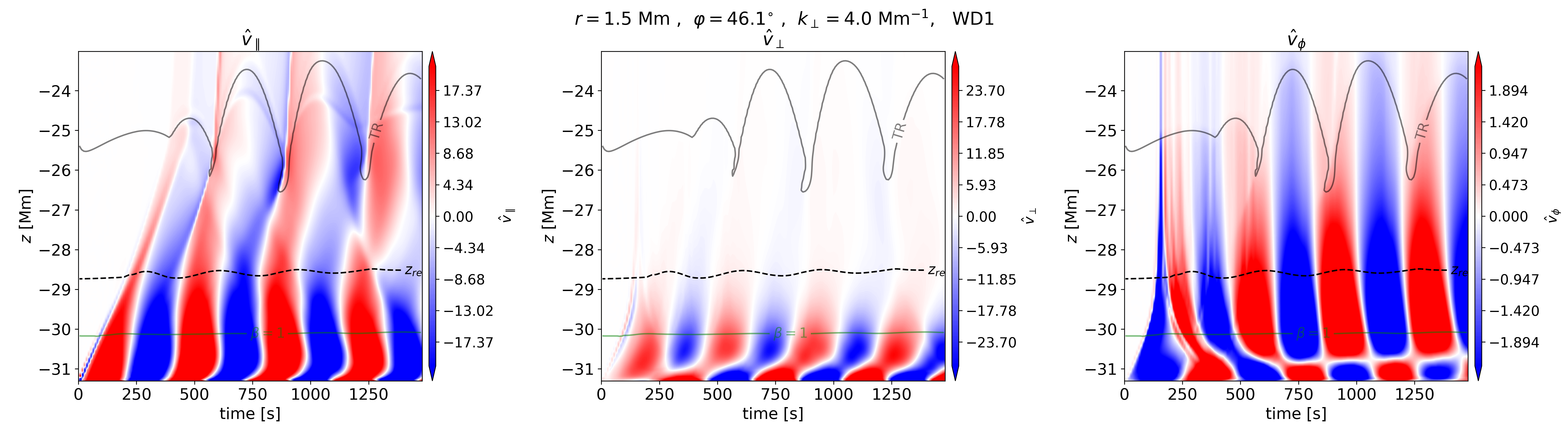}
    \caption{}
\end{subfigure}
\begin{subfigure}{0.97\textwidth}
    \includegraphics[width=\textwidth]{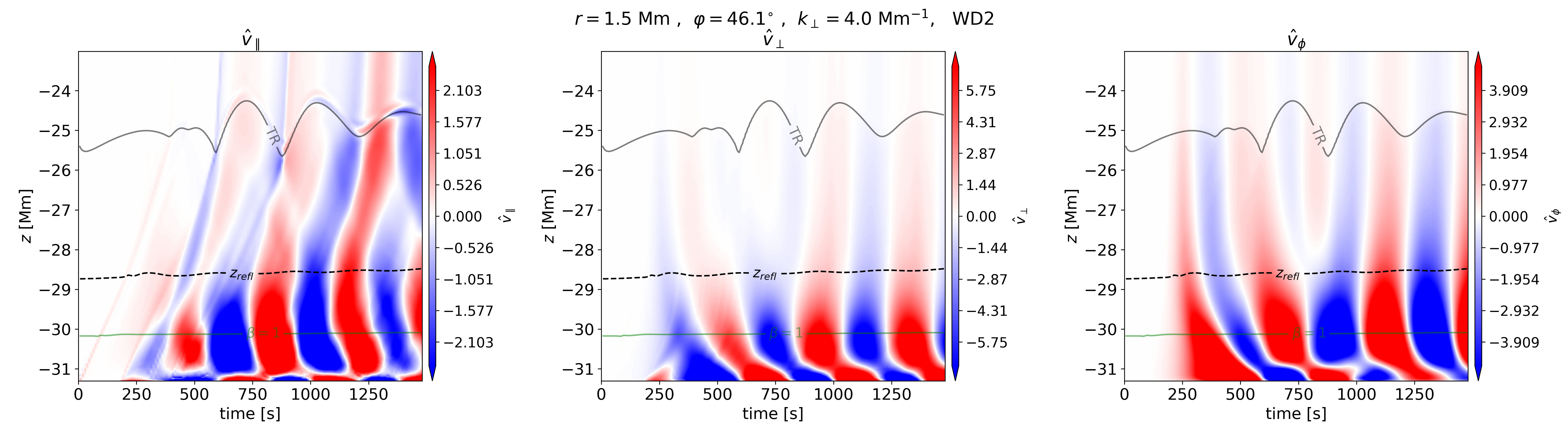}
    \caption{}
\end{subfigure}
\begin{subfigure}{0.97\textwidth}
    \includegraphics[width=\textwidth]{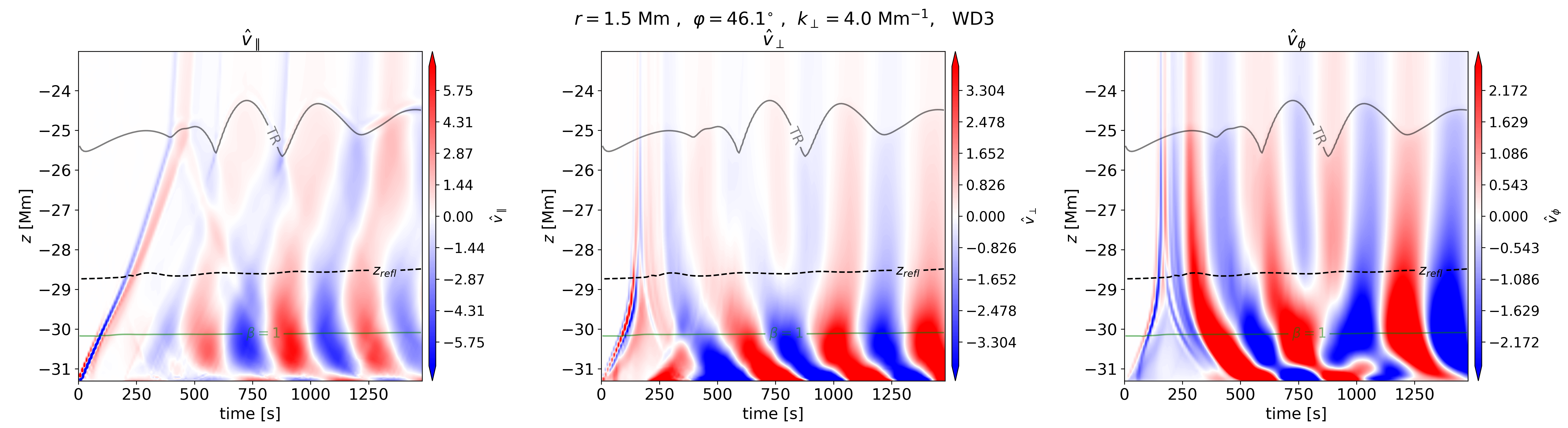}    
    \caption{}
\end{subfigure}
\caption{Time--distance diagrams of the scaled velocities for the slow magnetoacoustic (left), fast magnetoacoustic (middle), and Alfv\'{e}n (right) modes for an azimuthal slice $\varphi=46.1^{\circ}$ at a radial cut of $r=1.5$ Mm. Shown are the cases for different wave drivers highlighted in Table \ref{tab:wave_drivers} for (a) centred inclined driver of $15^{\circ}$ (WD1), (b) a shifted vertical driver (WD2) and (c) a shifted inclined wave driver of $20^{\circ}$ towards the loop (WD3). Otherwise, all markers in the plots are the same as Fig. \ref{fig:scaled_vels_15deg}.}
\label{fig:scaled_vels_driver_comparison}
\end{figure*}
The resulting time--distance diagrams of the velocity perturbations are shown for $r = 1.5$ Mm (i.e. inside the loop) in Fig. \ref{fig:scaled_vels_driver_comparison} for all wave drivers. The similarities between each of the subplots in Fig. \ref{fig:scaled_vels_driver_comparison} is especially striking. Namely, regardless of where the wave driver is located at the base of the photosphere, if there is strong transverse structuring present, the structure can be very efficient at absorbing the externally driven waves. The presence of the Alfv\'{e}n wave inside the loop for all wave drivers is a result of the transverse structuring, which enables the `loop' to absorb the fast magnetoacoustic wave component through `Fourier scattering' \citep{Cally2017, Khomenko2019, cally2019} and convert, through reflections due to the transverse gradients, to tube waves in the form of kink motions \citep[see e.g.][]{Riedl2019, Riedl2021, Skirvin2023ApJ}. This is an important result because it suggests that inhomogeneities in the direction transverse to the magnetic field allow significantly more energy to be channelled into the solar chromosphere and corona through the process of mode conversion, which has recently been discussed in \citet{Khomenko2019}.

The velocities depicting the fast magnetoacoustic and Alfv\'{e}n modes display similar characteristics, regardless of the position or inclination of the wave driver. This suggests that global motions are excited within the loop, which are channelled by the transverse structuring through reflections \citep{DeMoortel2012} and can explain the Alfv\'{e}nic motions reported by \citet{Skirvin2023ApJ}. The coupling between the fast magnetoacoustic mode and the Alfv\'{e}n mode can be seen by the similarities in the wave fronts of the velocities of both $\hat{v}_{\perp}$ and $\hat{v}_{\phi}$ in Fig. \ref{fig:scaled_vels_driver_comparison} and solidifies the consensus that such global transverse motions of a magnetic flux tube in a non-uniform plasma display characteristics of both fast magnetoacoustic and Alfv\'{e}n waves \citep{goo2009}.

The right hand side panels of Figs. \ref{fig:scaled_vels_15deg} and \ref{fig:scaled_vels_driver_comparison} show ridges with inclination suggesting downward propagation.
We have already suggested that this is a result of wave reflection from the transition region; however, the downward Alfv\'{e}n wave propagation could also be a consequence of the mode conversion from fast to Alfv\'{e}n waves resulting in downward propagating (converted) Alfv\'{e}n waves. This is a result of the inclination angle between the driven wave vector and the magnetic field vector \citep{Cally2008}. For example, the right hand side panels of Fig. \ref{fig:scaled_vels_driver_comparison} display the time--distance plots when the wave driver is positioned outside of the loop structuring. Therefore, the horizontally propagating wave fronts create a greater attack angle with the magnetic field vector within the conversion region inside the loop, and may form downward propagating Alfv\'{e}n waves in a process that is outlined in \citet{Cally2008} and \citet{Khomenko2012}.

\subsection{Wave energy fluxes}\label{subsec:wave_fluxes}

The total wave energy flux can be written as follows \citep{Riedl2021}:
\begin{equation}\label{eqn_flux}
    F = -\frac{1}{\mu_0}\left(\mathbf{v} \times \mathbf{B}\right) \times \mathbf{B} + \left( \frac{\rho v^2}{2} + \rho \Phi + \frac{\gamma}{\gamma+1}p \right)\mathbf{v}.
\end{equation}
The first term on the right hand side of Eq. (\ref{eqn_flux}) corresponds to the Poynting flux (i.e. the magnetic flux), whereas the other term describes the hydrodynamic (HD) component of the total energy flux with contributions from kinetic, gravitational and thermal components. Providing a detailed analysis regarding the energy carried by slow, fast, and Alfv\'{e}n waves in the simulation is extremely challenging. This is a result of the non-uniformity of the model, a consequence of both vertical and transverse structuring. Therefore, identifying signatures of pure slow, fast, and Alfv\'{e}n waves in the simulation is actually impossible since the pure MHD modes do not exist, rather they are coupled with one another and display mixed properties \citep[see e.g.][]{goo2019}. One approach for determining the energy associated with each wave mode could be to analyse the energy contained within the acoustic (inside the HD component) and magnetic wave energy fluxes contained in Eq. (\ref{eqn_flux}). However, this approach only separates the slow magnetoacoustic mode, which is predominantly acoustic in a low beta plasma. The magnetic wave energy flux would still contain contributions from both fast magnetoacoustic and Alfv\'{e}n waves.

Therefore, to provide an estimate of the contribution of individual MHD modes, we determined the `potential' energy flux contained in the three different modes as follows \citep[e.g.][]{Felipe2012, Yadav2022}:
\begin{eqnarray}
    \label{linear_parallel_energy}
    F_{\parallel} &=& \rho v_{\parallel}^2 c_s , \\
    \label{linear_perp_energy}
    F_{\perp} &=& \rho v_{\perp}^2 \sqrt{c_s^2 + v_A^2} , \\
    \label{linear_azimuthal_energy}
    F_{\phi} &=& \rho v_{\phi}^2 v_A ,    
\end{eqnarray}
where these expressions combine the ram pressure in the decomposed wave modes from Eqs. (\ref{decomp_parallel})-(\ref{decomp_perp}) with the relevant group speed of the mode in question. For a low-$\beta$ plasma, the group speed is equal to the sound speed for the slow magnetoacoustic mode, the Alfv\'{e}n speed for the Alfv\'{e}n mode, and the combined sound-Alfv\'{e}n speed (fast speed) for the fast magnetoacoustic mode. As a result, a discussion on the potential wave energy fluxes described by Eqs. (\ref{linear_parallel_energy})-(\ref{linear_azimuthal_energy}) are only physically valid above the equipartition layer. The following discussion will compare the different wave drivers to study the relationship between the transverse structuring and mode conversion in terms of the potential wave energy flux.

\begin{figure*}
    \centering    \includegraphics[width=0.99\textwidth]{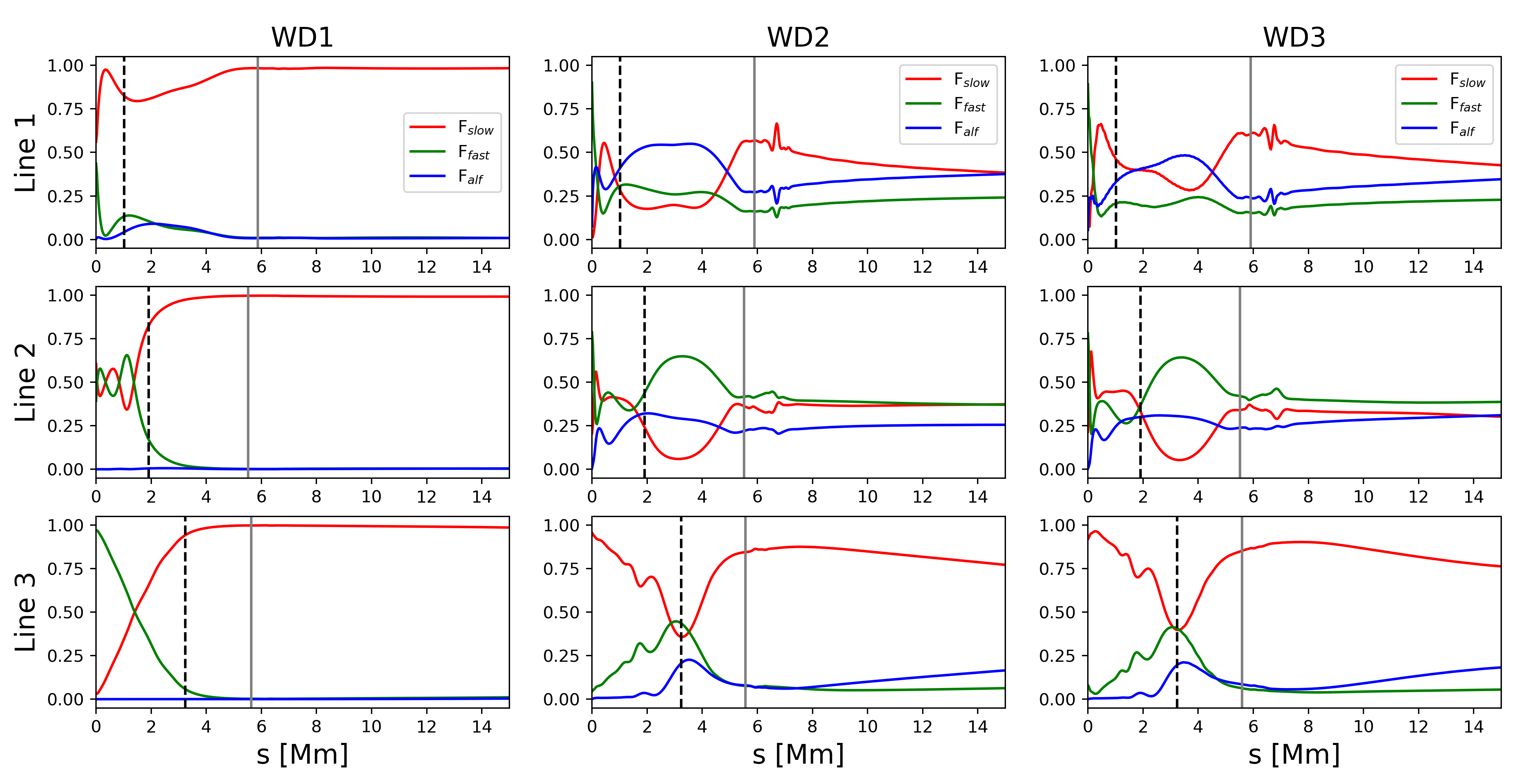}
    \caption{Potential wave energy fluxes calculated using Eqs. (\ref{linear_parallel_energy})-(\ref{linear_azimuthal_energy}).\ We show the `available' energy flux associated with slow magnetoacoustic (red curve), fast magnetoacoustic (green curve), and Alfv\'{e}n (blue curve) waves. The energy fluxes are plotted along the respective field lines with the distance along the field line denoted by the variable $s$. The energy fluxes are plotted for the three different field lines shown in Fig. \ref{fig:Bfield_strength}, with line $1$ (top row), line $2$ (middle row) and line $3$ (bottom row). The left column shows the results for the simulation of the centred, inclined wave driver (WD1), the middle column shows the result for the shifted vertical wave driver (WD2) whereas the right column is for WD3 (Table \ref{tab:wave_drivers}). Finally, the average position of the plasma-$\beta=1$ layer and transition region are denoted by the black dashed line and grey solid line, respectively. All plots are normalised such that the sum of all energy fluxes equals unity at every respective height. }
    \label{fig:potential_wave_fluxes}
\end{figure*}

Figure \ref{fig:potential_wave_fluxes} plots the normalised potential wave energy fluxes calculated using Eqs. (\ref{linear_parallel_energy})-(\ref{linear_azimuthal_energy}) as a function of the distance along selected field lines \citep[see e.g.][]{Yadav2022}. These expressions are averaged over all azimuthal angles and additionally averaged throughout the four driving periods in the simulation. The potential wave energy fluxes are calculated along the three field lines shown in Fig. \ref{fig:Bfield_strength}. We begin by discussing the left hand panels of Fig. \ref{fig:potential_wave_fluxes}, which highlight the potential wave energy fluxes for WD1 and correspond to the study by \citet{Skirvin2023ApJ}. On field lines $2$ and $3$ there is very little potential wave energy flux associated with fast and Alfv\'{e}n waves, whereas almost all of the wave energy flux is embedded within slow magnetoacoustic modes above the equipartition layer. This result can be understood by realising the inclined driver excites waves that are predominantly aligned with the magnetic field and, at these locations outside of the loop structuring, the primary velocity perturbations remain aligned with the magnetic field. On field line $1$, which is located in a region of stronger transverse gradients and also within the position of the inclined driver, we also observe a similar distribution of the potential wave energy fluxes. Slightly above the equipartition layer there is a small increase in fast mode energy flux, which is compensated for by a reduction of slow mode energy flux. At heights above the slow to fast conversion, there is a small increase in the potential wave energy flux associated with Alfv\'{e}n waves, resulting from fast to Alfv\'{e}n mode conversion in our stratified 3D domain. However, we can conclude that waves driven within the loop structuring possess energy fluxes that can be mainly attributed to slow modes. Nonetheless, by breaking the azimuthal symmetry and considering an inclined driver, small amounts of energy flux are associated with the fast and Alfv\'{e}n modes, resulting from mode conversion processes in the stratified atmosphere.

In addition, the bottom panel of Fig. \ref{fig:potential_wave_fluxes} shows the potential wave energy fluxes along field line $3$, which is rooted at the centre of the locally applied (shifted) driver and outside of the loop structuring. For both shifted wave drivers WD2 (vertical) and WD3 (inclined) we observe a very similar behaviour of each of the potential wave energy fluxes associated with slow, fast, and Alfv\'{e}n modes on this field line. In the vicinity of the equipartition layer, denoted by the vertical dashed line in Fig. \ref{fig:potential_wave_fluxes}, the incoming fast acoustic wave splits into slow and fast magnetoacoustic modes. Slightly above the equipartition layer, where the fast mode reflection height is located, we see a peak in the amplitude of the potential flux of the Alfv\'{e}n mode, which is expected from double-mode conversion in a $3$D domain. At heights greater than the fast mode reflection height, the potential wave energy flux associated with fast waves quickly decays and becomes evanescent, consistent with the study of \citet{Khomenko2012}. Above the transition region, given by the solid grey line in Fig. \ref{fig:potential_wave_fluxes}, the majority of the potential wave energy flux is contained within slow modes on field line $3$, which is expected due to the nature of the driver, and a small contribution from the Alfv\'{e}n mode is present resulting from double-mode conversion. The behaviour observed on field line 3 shares striking characteristics of the well-known (double) mode conversion shown to occur in thick flux tube sunspot atmospheres \citep{Khomenko2012, Felipe2012}. This can be understood by recognising that, at this location positioned away from the loop structuring, the transverse gradients are weak. Hence, the plasma can be considered as locally uniform in the radial (transverse) direction and the mode conversion processes behave similarly to previous studies considering a thick flux tube with varying transverse structuring.

Turning attention now to the middle panels of Fig. \ref{fig:potential_wave_fluxes}, which highlight the potential wave energy fluxes on field line $2$, where the transverse structuring provided through the magnetic field becomes more important. One immediately notices a clear difference between the potential wave energy fluxes on field line $2$ compared with field line $3$: the energy flux associated with the fast magnetoacoustic mode dominates in the chromosphere $2$ Mm $< s < $ $5$ Mm. Providing an exact explanation as to why this occurs is very difficult as there may be multiple factors each playing a role. On one hand, due to the greater attack angle between the wavevector of the driven waves and the magnetic field at the equipartition region, the incoming fast acoustic waves may preferentially convert to fast magnetoacoustic modes over slow magnetoacoustic modes. Moreover, the tube structuring provided by the magnetic field may also manifest inside the potential fast magnetoacoustic wave energy flux as a result of tube waves, such as sausage and kink modes \citep{Riedl2019, Riedl2021, Skirvin2023ApJ}, which produce velocity perturbations normal to the magnetic field. This occurs due to fast magnetoacoustic wave refraction, reflection and scattering resulting from the stronger transverse gradients across the magnetic field \citep{DeMoortel2012, Cally2017, Khomenko2019}. Furthermore, looking at the potential wave energy flux of the Alfv\'{e}n mode on field line $2$, it is no longer clear that this energy is a result of the linear double-mode conversion mechanism, as the potential energy flux appears to be relatively constant throughout the chromosphere. In the corona on field line $2$, there is a balanced contribution from the slow magnetoacoustic, fast magnetoacoustic and Alfv\'{e}n modes to the total potential wave energy flux, highlighting the coupled nature of MHD waves in a non-uniform plasma. 

\begin{figure*}
    \centering    \includegraphics[width=0.99\textwidth]{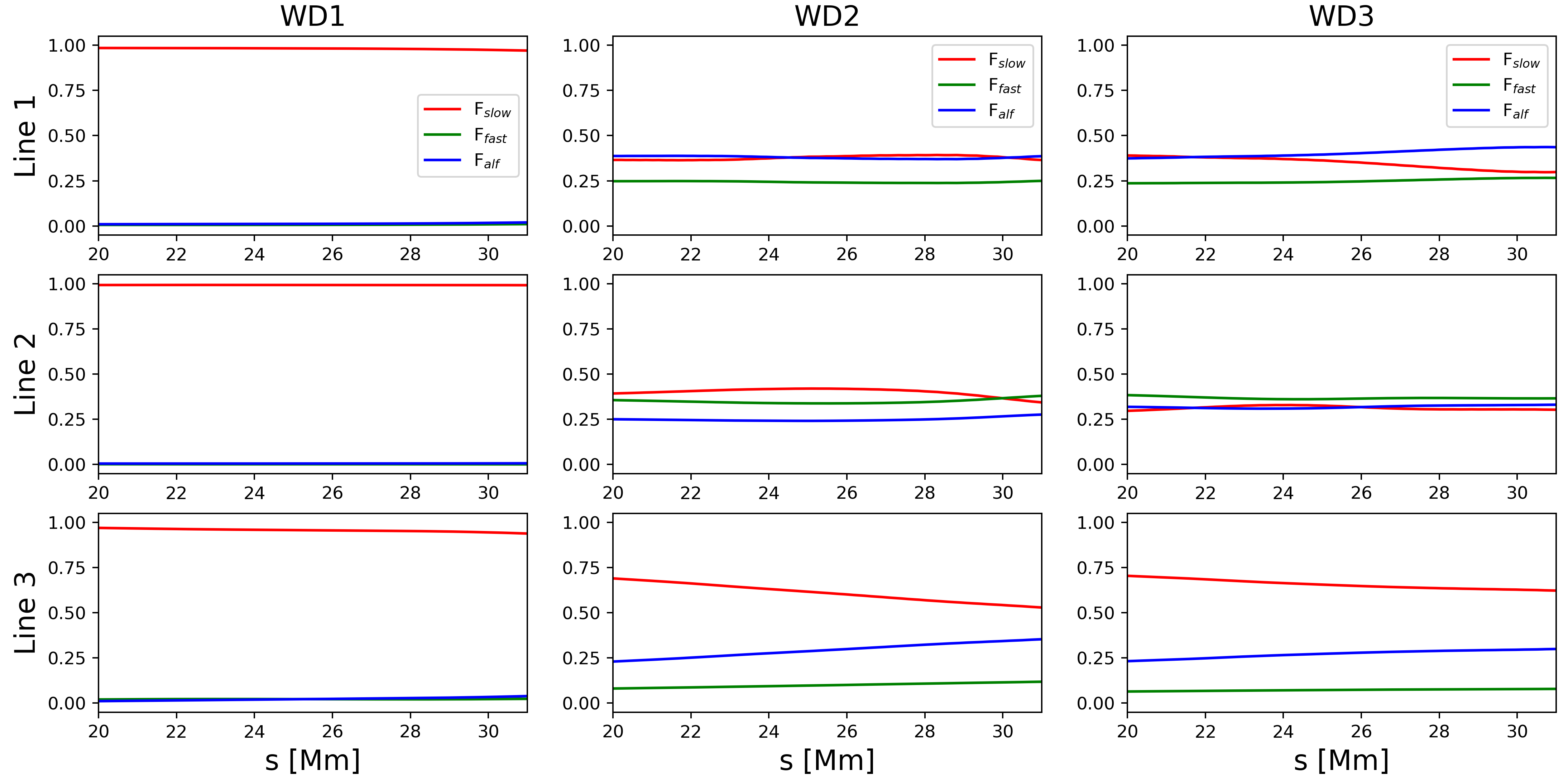}
    \caption{Same as Fig. \ref{fig:potential_wave_fluxes} but over greater heights, including the apex of the loop at $s = 31$ Mm.}
    \label{fig:potential_wave_fluxes_APEX}
\end{figure*}

On field line $1$, the magnitude of the potential energy flux associated with Alfv\'{e}n waves dominates in the chromosphere, as seen in the middle and right panels on the top row of Fig. \ref{fig:potential_wave_fluxes}. In the corona, the Alfv\'{e}n wave potential energy flux becomes dominant, and, at the apex of the loop around $s=30$ Mm (shown in Fig. \ref{fig:potential_wave_fluxes_APEX}) there is greater wave flux associated with Alfv\'{e}n waves than slow magnetoacoustic waves. Overall, the potential wave energy flux associated with Alfv\'{e}n waves contributes roughly $50$\% to the total wave energy flux on field line $1$ at the apex of the loop for our shifted wave drivers. This can be understood by the presence of the Alfv\'{e}nic motions reported by \citet{Skirvin2023ApJ}, which can be thought of as surface Alfv\'{e}n waves existing on individual field lines throughout the non-uniform plasma, providing a greater contribution to $\hat{v}_{\varphi}$ across all azimuthal angles as a result of the global motion supported by the transverse structuring. There is clearly a strong effect of the potential Alfv\'{e}n wave energy flux on the transverse structuring in the model provided by the magnetic field as shown in Fig. \ref{fig:potential_wave_fluxes}. This can immediately be seen by comparing the blue curves in each row. On field line $1$ (top panels) the potential wave energy flux associated with Alfv\'{e}n modes exceeds that of slow magnetoacoustic modes at the apex in the corona (see Fig. \ref{fig:potential_wave_fluxes_APEX}) even when the wave driver is positioned away from the loop structuring. However, where the effect of structuring is weaker (bottom panels), the potential wave energy flux in Alfv\'{e}n modes, resulting from double-mode conversion, is significantly smaller than that associated with slow magnetoacoustic modes, regardless of the fact that the wave driver is rooted in the field line used for tracing here. Overall, there does not appear to be any major differences from varying the inclination of the wave driver on the potential wave energy fluxes on each of the field lines considered for analysis here.

\subsection{Wave energy flux absorption via transverse structuring}
To study the effectiveness of the loop structuring at absorbing the driven wave energy flux, we integrated the wave energy flux provided in Eq. (\ref{eqn_flux}) over the final driving period in all simulations and further integrated throughout all azimuthal angles. We integrated over the final driving period only because the simulation had reached a steady state by this point. The loop model considered in this work does not have a clear boundary denoted with a discontinuity in a plasma variable, such as density or magnetic field; rather, the plasma is continuous in the transverse direction. Therefore, we used field line $2$ (see Fig. \ref{fig:Bfield_strength}) as a proxy to define the `boundary' of the loop and integrated the fluxes at each height, both inside and outside the location of field line $2$.
\begin{figure*}
    \centering    \includegraphics[width=0.99\textwidth]{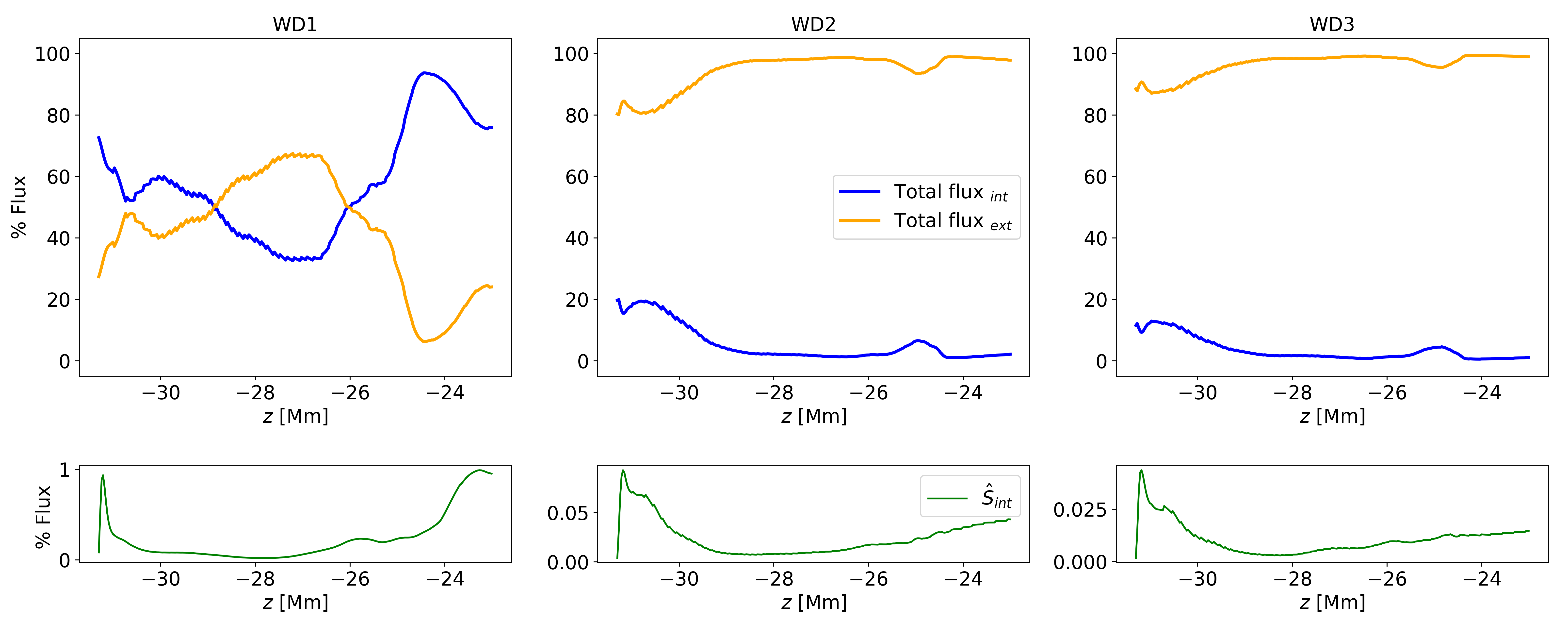}
    \caption{Wave energy flux percentages, calculated using Eq. (\ref{eqn_flux}), contained within the loop (blue line) and outside the loop (orange line) at different heights for the wave drivers modelled in this work. The loop boundary is defined as the local radial position of field line 2 at each height. The wave energy flux is integrated over the final driving period (T3-T4) and additionally integrated over all azimuthal angles. The bottom panels display the contribution of the total Poynting flux inside the loop (green line). The left, middle and right panels indicate the simulations of WD1, WD2, and WD3, respectively. }
    \label{fig:loop_fluxes_lower_atms}
\end{figure*}

The resulting percentages of the total fluxes (HD + Poynting) both inside and outside the loop for all wave drivers are displayed in Fig. \ref{fig:loop_fluxes_lower_atms} for the lower solar atmosphere in the model, ranging from the photosphere, where the waves are driven, to slightly above the transition region. Unsurprisingly, most of the total flux at each height is located inside the loop for the simulation of WD1, where the driver is centred at the loop axis and located inside the loop. Moreover, the HD component of the wave energy flux dominates the total flux inside the loop when compared to the Poynting flux (bottom panels of Fig. \ref{fig:loop_fluxes_lower_atms}), although this is expected due to the nature of the acoustic wave driver. In the Chromosphere, there appears to be significant energy flux leakage as the total flux at these heights is greater at radial positions greater than the local position of field line $2$. The lateral leakage of wave energy flux in the lower solar atmosphere was previously reported by \citet{Riedl2021}. However, in the corona above the transition region, at heights roughly $z>-24$ Mm, the total flux inside the loop dominates for WD1. This analysis for WD1 suggests that wave energy leakage can be expected to be ubiquitous in the chromosphere for waveguide structures with continuous transverse gradients. Although structures with more obvious boundaries, for example the discontinuous profiles considered in classic flux tube models \citep[e.g.][]{edrob1983}, may be able to trap more energy flux inside.

Turning attention now to the simulations where the wave driver is positioned outside of the loop structuring, the middle and right panels of Fig. \ref{fig:loop_fluxes_lower_atms}, we observe that the majority of the total flux in the domain is located outside of the loop, with only a few percent at best of the total flux being absorbed by the loop structuring in the lower atmospheric domain of the model. Similarly to the case of WD1, essentially all of the flux inside the loop structuring is acoustic in nature, with less than $0.1 \%$ of the Poynting flux contributing. The wave energy flux absorption is similar for both externally driven scenarios WD2 and WD3; however, it appears that slightly more wave energy flux (a few percent) is absorbed by the loop when the wave driver is vertical (WD2) as opposed to the case when the wave driver is inclined towards the loop (WD3). This may be a result of greater wave reflection by the transverse structuring when the waves are driven at an angle towards the loop due to the stronger transverse gradients that the waves encounter.

The preceding discussion has focused on the efficiency of transverse structuring at absorbing wave energy flux in the lower solar atmosphere, where the majority of the domain is described with a plasma-$\beta$ greater than unity, such that the thermodynamic properties of the plasma dominates over the magnetic field. Figure \ref{fig:loop_fluxes_lower_atms} suggests that at heights above the transition region, the percentage of wave energy flux contained within the loop structuring begins to increase; therefore, it may be instructive to conduct a similar analysis at the loop apex, where the plasma-$\beta$ is much smaller than unity and the loop structuring controls the overall plasma dynamics.
\begin{figure}
    \centering    \includegraphics[width=0.45\textwidth]{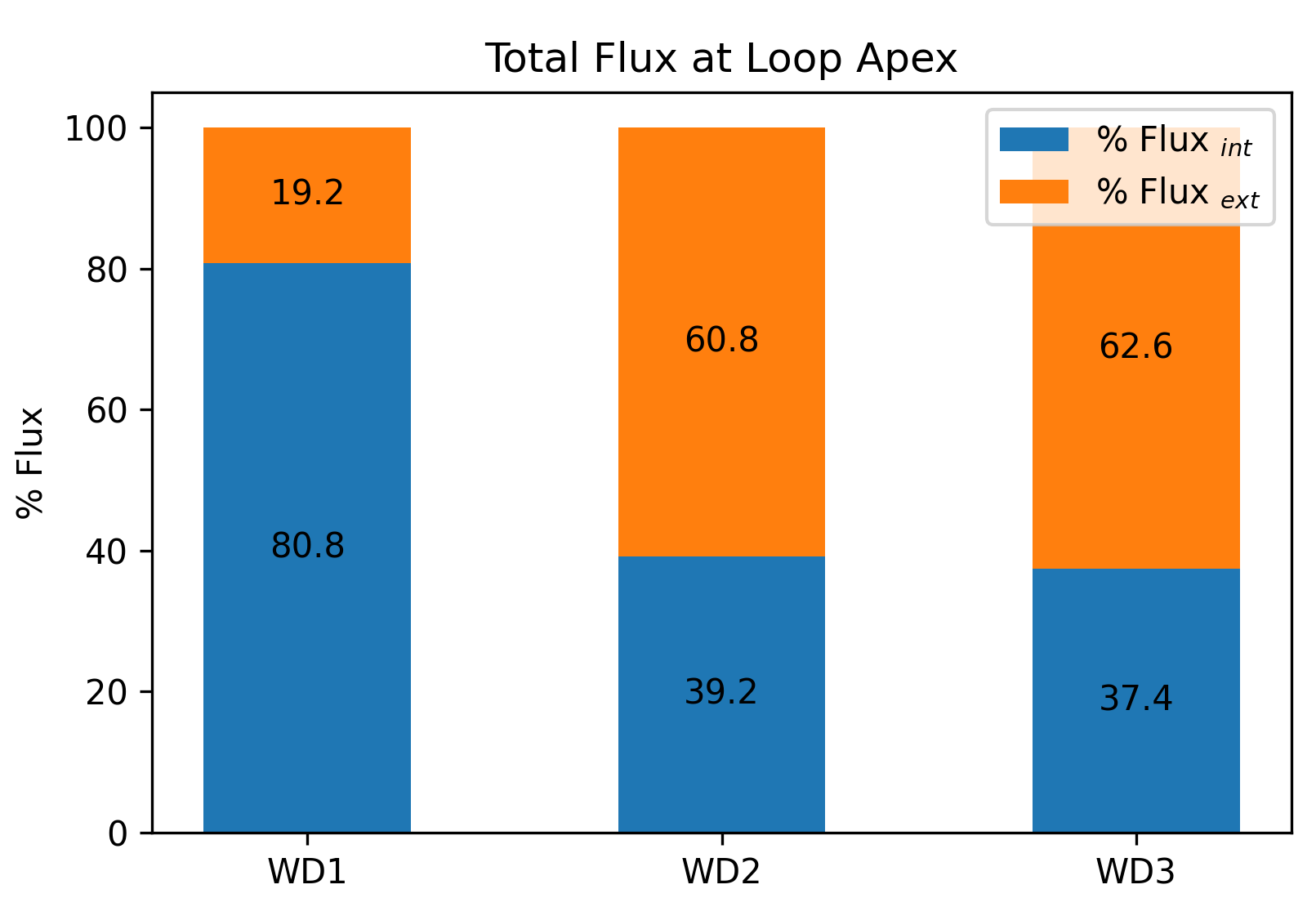}
    \caption{Percentage distribution of wave energy fluxes, calculated using Eq. (\ref{eqn_flux}), both inside (blue) and outside (orange) the loop structure at the loop apex ($z=0$ Mm or equivalently a height of $31$ Mm above the photosphere) for the three wave drivers considered in this study.}
    \label{fig:loop_fluxes_apex}
\end{figure}
Figure \ref{fig:loop_fluxes_apex} shows the percentage distribution of wave energy flux, calculated using Eq. (\ref{eqn_flux}), both inside and outside the loop structuring; it is the same as Fig. \ref{fig:loop_fluxes_lower_atms} except the loop apex ($z=0$ Mm) is located $31.31$ Mm above the photosphere. We can see from Fig. \ref{fig:loop_fluxes_apex} that a significant proportion of the total wave energy flux is now concentrated within the loop structuring at this height. In fact, even for the scenario where the wave driver is positioned outside of the loop, over a third of the total energy flux becomes channelled by the loop at the apex. Moreover, at the base of the domain, $100$\% of the wave energy flux is initially located outside the loop for the case of the shifted wave drivers (WD2 and WD3); however, at the apex of the loop in the corona, nearly $40$\% of this wave energy flux has now been absorbed by the loop. It is important to note that because of the shape of the driver, it produces horizontally expanding wave fronts, in addition to vertically propagating waves. Some of these horizontally travelling wave fronts move towards the flux tube centre in the lower atmosphere and eventually appear at higher heights encapsulated in Fig. \ref{fig:loop_fluxes_apex}, which displays the sum of the wave flux absorbed by both the horizontally expanding wave fronts and the vertically propagating waves. This is a significant result as it highlights the efficiency of coronal loops at trapping and transporting wave energy flux throughout the solar corona and provides evidence that coronal loops play the role of energy ducts. Moreover, this is direct evidence that transverse structuring plays a pivotal role in channelling waves into the solar corona. It must be stressed that the total physical energy flux reaching the corona is larger for the case of WD1 when compared with WD2 and WD3; however, the energy flux in this case is associated with slow modes. On the other hand, when the wave driver is positioned outside the loop structuring, although the total physical energy flux contained within the loop may be comparably less than that of a foot-point driver, the wave energy flux inside the loop is now equally divided between all types of MHD modes.

\section{Conclusions}\label{sec:conclusions}

In this work we have studied the mode conversion and wave energy flux absorption processes in a gravitationally stratified and transversely structured solar atmosphere. The well-known linear mode conversion processes between slow, fast, and Alfv\'{e}n waves at the equipartition layer are present; however, there is also an additional effect due to the transverse structuring. The loop-like structure, resulting from the transverse magnetic field gradient, behaves as a conduit for Alfv\'{e}n waves into the magnetically dominated solar corona. The subsequent absorption and refraction of the fast magnetoacoustic waves by the strong transverse gradients cause them to manifest as tube waves of a magnetic cylinder, which can explain the results of, for example, \citet{Riedl2021} and \citet{Skirvin2023ApJ}. In this scenario, the initially acoustic waves in the high plasma-$\beta$ photosphere develop magnetic properties once they pass the equipartition layer, where $c_s=v_A$. The radial structuring, provided by the magnetic field, then permits these waves, which have now developed magnetic properties, to convert to tube waves of a magnetic cylinder through Fourier scattering \citep[e.g.][]{Khomenko2019, cally2019}; they are then guided by the field through the solar atmosphere. We have also demonstrated that regardless of the location of the acoustic wave driver in the photosphere, the loop structure still absorbs and channels waves upwards through the stratified atmosphere. 

Moreover, in regions where the transverse gradients are strong, we find that there is significantly less reflection of Alfv\'{e}n waves at the transition region, similar to the results of \citet{Khomenko2019}, whereas in regions where the structuring is weak, wave reflection at the transition region is much greater. This is complimented by a greater percentage of Alfv\'{e}n wave energy flux in the corona along field lines.\ Here the transverse gradients are strong compared to regions of weak structuring, where the majority of the total wave energy flux in the corona is associated with the slow magnetoacoustic mode due to the nature of the wave drivers employed in this work. In regions where the transverse structuring is weak, we observe characteristics very similar to those of the classic double-mode conversion \cite[e.g.][]{Khomenko2012, Felipe2012} upon evaluation of the potential wave energy fluxes.

Furthermore, we examined the efficiency of loop-like structuring at absorbing external wave energy flux by investigating the associated wave energy fluxes inside and outside the loop from acoustic wave drivers, which are located outside the loop structure at the photospheric base of the domain. In the lower solar atmosphere, such as the photosphere and chromosphere of the model, the majority of the wave energy flux is located outside of the loop structuring, with wave leakage a ubiquitous phenomenon. This is even true for the case when the wave driver is positioned inside the loop (WD1). This may be a result of the local thermodynamic properties of the plasma in the lower solar atmosphere, where the plasma-$\beta$ is not much lower than unity. For the cases of the shifted wave drivers, WD2 and WD3, initially all of the wave energy flux is outside the loop; however, we find that at the apex of the loop in the magnetically dominated corona, over a third of the total wave energy flux at this height is now located within the loop, suggesting that coronal loops do indeed play a very important role in channelling energy within the corona. Upon comparison of the external wave drivers positioned outside the loop structuring in this work, we find that slightly less wave energy flux is absorbed by the loop when the wave driver is inclined towards the loop. An explanation for this could be that the waves that are driven at an angle towards the loop structuring must propagate across stronger transverse gradients, as opposed to purely vertical driven waves, which results in greater wave reflection and, ultimately, reduced absorption rates. 

The results presented in this study have potentially tantalising implications for explaining oscillatory phenomena observed in the solar atmosphere. For example, decayless oscillations of coronal loops are usually considered to be driven either within the loop structure or from directly below the footpoints \citep[e.g.][]{Guo2019, Karampelas2019, Afan2019, Gao2023}. However, we have provided evidence that such oscillations may also be produced by a Fourier scattering effect \citep{cally2019,Khomenko2019} from external waves generated by, for example, wave-generating magnetic reconnection processes \citep{McLaughlin_et_al_2012} that are absorbed by the loop or generated by linear mode conversions \citep{Schunker2006, Khomenko2012} from acoustic perturbations at the photosphere. Furthermore, the physical processes of mode conversion discussed in this work contribute to the understanding of the associated periodicity of such oscillations matching those of the acoustic modes of the solar interior.

For all wave drivers adopted in this study, there are clear signatures in the simulations that the driven acoustic waves interact non-linearly with the transition region (e.g. Figs. \ref{fig:scaled_vels_15deg} and \ref{fig:scaled_vels_driver_comparison}) to produce periodic jet-like features \citep[e.g.][]{scu2011, Skirvin2023_JET}, which may not only act as wave guides themselves to transport energy to the solar corona, but also appear to display similar characteristics of propagating disturbances \citep{Samanta2015}. Conducting an in depth study of the waves' interaction with the transition region is the focus of an upcoming study, as these features can be detected in observations and used to conduct seismology studies of the local plasma and thus provide a greater understanding of the driving mechanism of solar jets in the lower solar atmosphere.

\begin{acknowledgements}
SJS and TVD were supported by the European Research Council (ERC) under the European Union's Horizon 2020 research and innovation programme (grant agreement No 724326) and the C1 grant TRACEspace of Internal Funds KU Leuven. The results received support from the FWO senior research project with number G088021N.
\end{acknowledgements}

\bibliography{ref}
\bibliographystyle{aa}

\begin{figure*}
\begin{subfigure}{0.5\textwidth}
    \includegraphics[width=\textwidth]{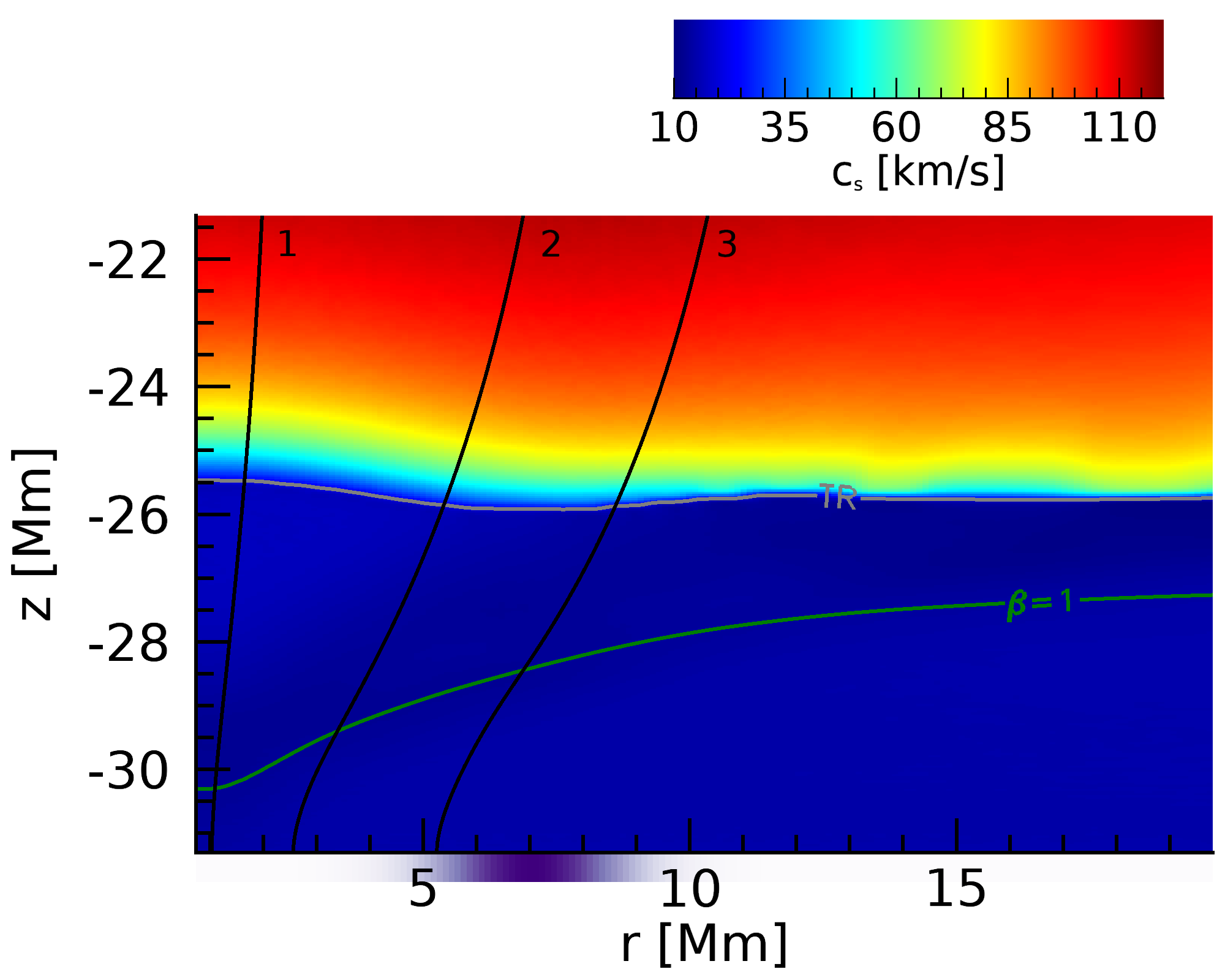}
    \caption{}
\end{subfigure}
\begin{subfigure}{0.5\textwidth}
    \includegraphics[width=\textwidth]{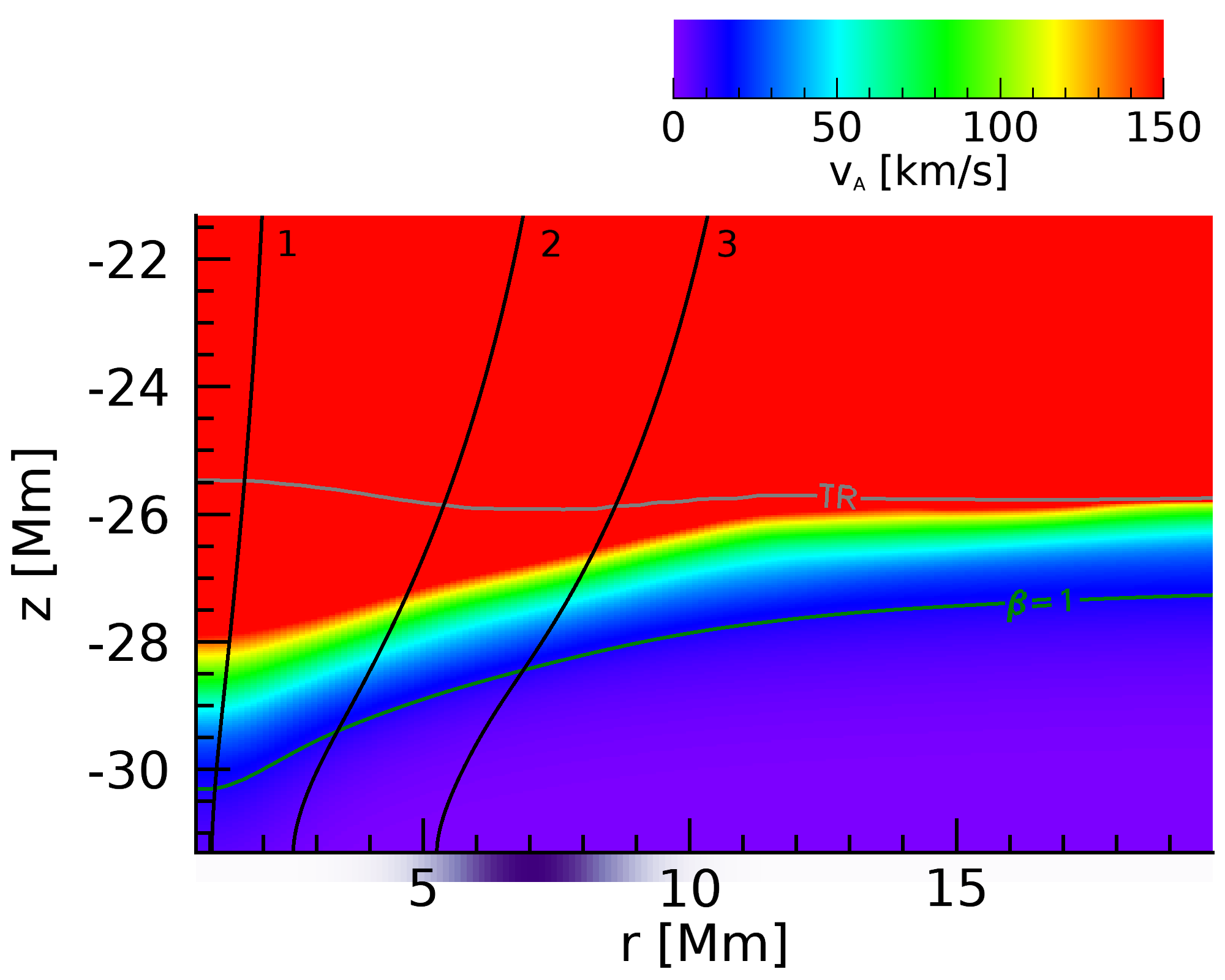}
    \caption{}
\end{subfigure}
\begin{subfigure}{0.5\textwidth}
    \includegraphics[width=\textwidth]{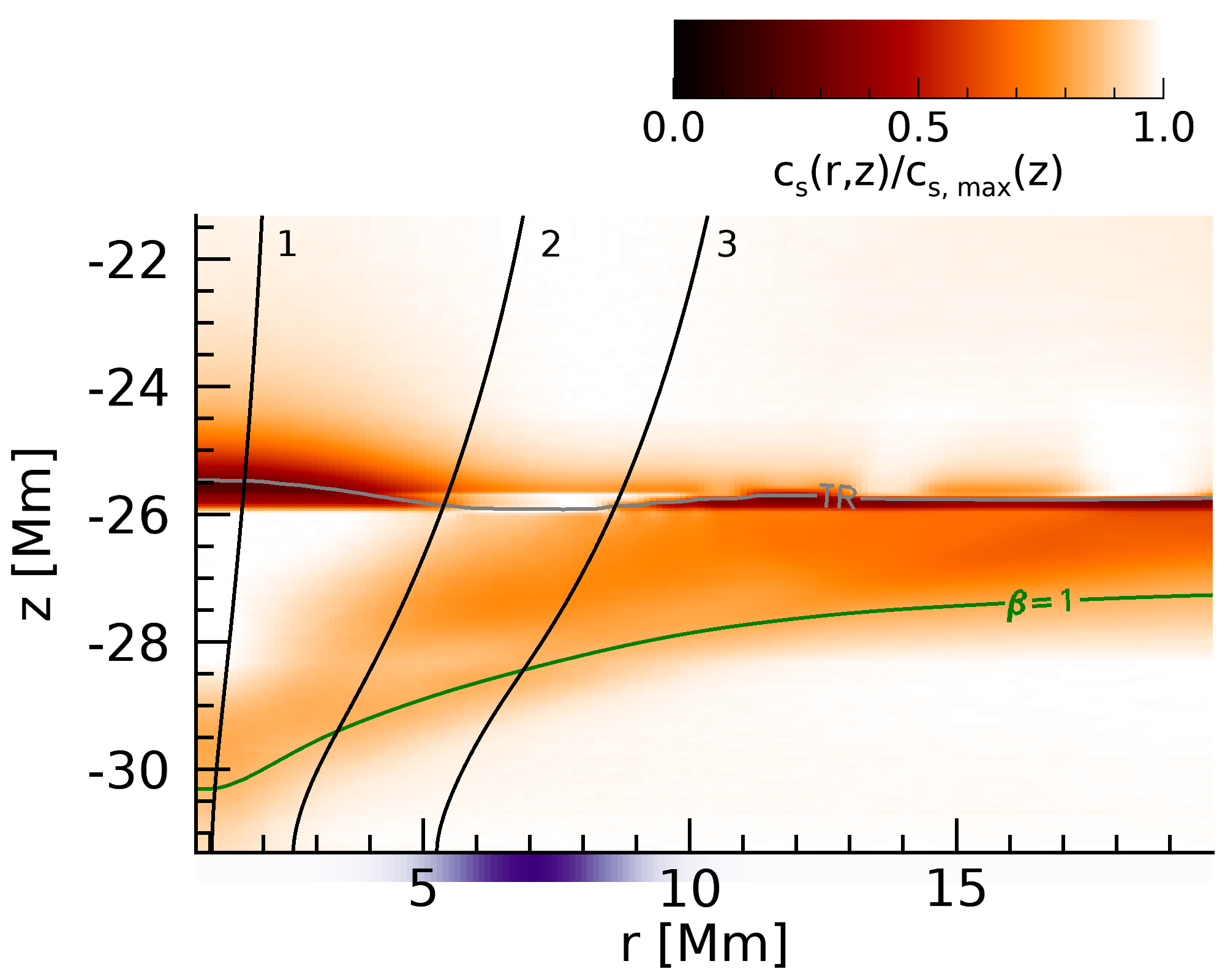}    
    \caption{}
\end{subfigure}
\begin{subfigure}{0.5\textwidth}
    \includegraphics[width=\textwidth]{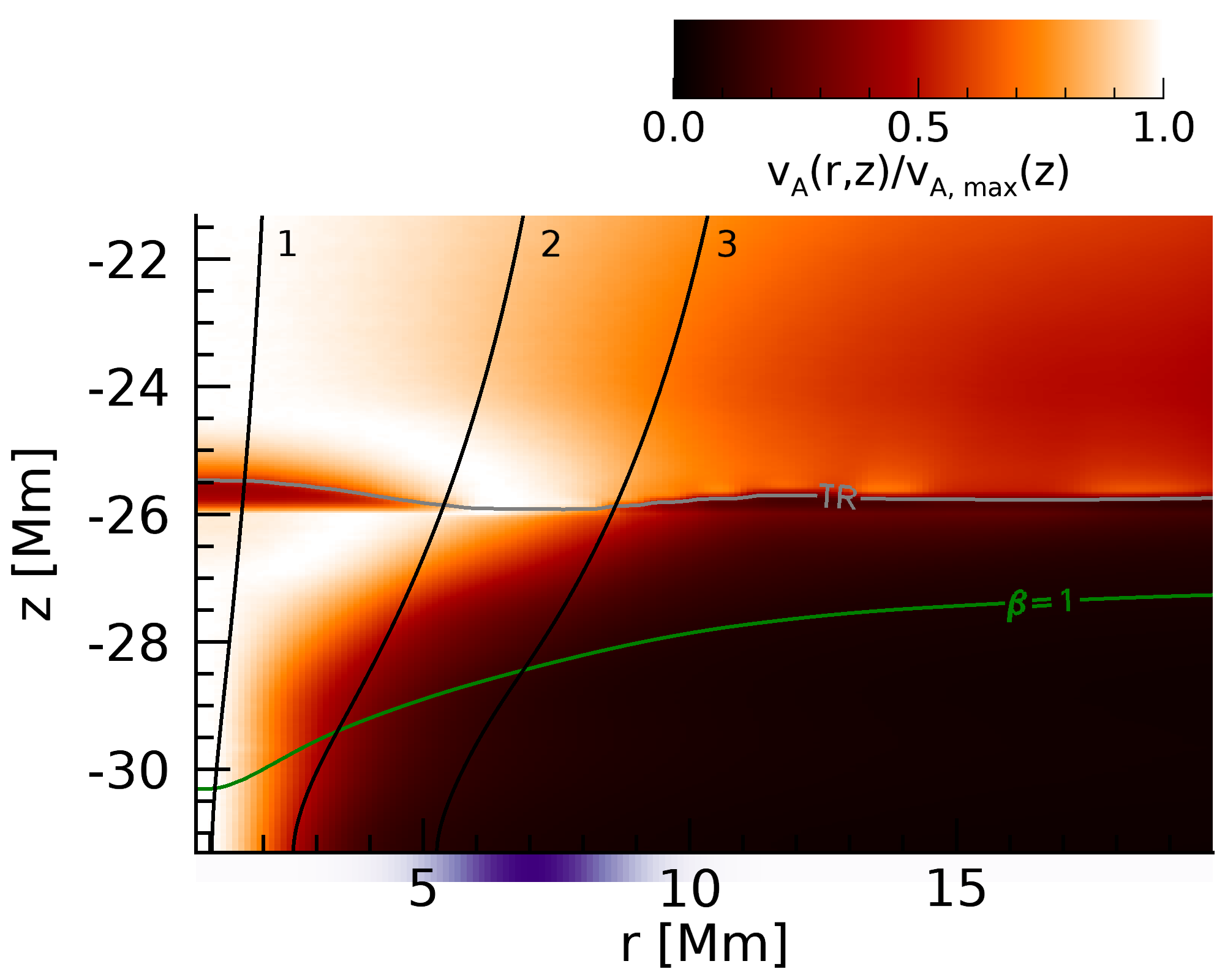}    
    \caption{}
\end{subfigure}
\caption{Panels (a) and (b) respectively show the characteristic sound speed ($c_s$) and Alfv\'{e}n speed ($v_A$) in the initial simulation domain at $t=0$ s. Panels (c) and (d) show the relative speeds with respect to their maximum value for all $z$, which highlights the horizontal structuring.}
\label{fig:characteristic_speeds}
\end{figure*}

\begin{appendix} 
In Fig. \ref{fig:characteristic_speeds} we display snapshots of the initial characteristic speeds -- Alfv\'{e}n ($v_A$) and acoustic ($c_s$) -- in the numerical domain to aid understanding of the parameters directly related to wave propagation. We also illustrate the horizontal gradients of these characteristic speeds by plotting the normalised speed as a function of the $r$-coordinate for all heights, $z$. It is normalised with respect to the maximum value of the respective characteristic speed at every height.

\end{appendix}

\end{document}